\documentclass[journal,draftcls,onecolumn,12pt,twoside]{IEEEtranTCOM}
\IEEEoverridecommandlockouts
\usepackage{amsmath,amssymb,amsfonts,amsthm}
\usepackage{dsfont}
\usepackage{algorithm}
\usepackage{algorithmic}
\usepackage{graphicx}
\usepackage{textcomp}
\usepackage{xcolor}
\usepackage{pgf}
\usepackage{tikz}
 \usetikzlibrary{arrows,automata}
\usepackage{verbatim}
\usepackage{subcaption}
\usepackage{url}
\usepackage{balance}
\usepackage{multirow}
\usepackage{setspace}
\usepackage{cite}

\theoremstyle{definition}
\newtheorem{theorem}{Theorem}

\newtheorem{lemma}{Lemma}
\newtheorem{assumption}{Assumption}

\begin{document}

\title{\Large Semantics-Aware Active Fault Detection in Status Updating Systems}

\author{\normalsize George Stamatakis,
	Nikolaos Pappas,
	Alexandros Fragkiadakis,
	and~Apostolos Traganitis
	\thanks{G. Stamatakis, A. Fragkiadakis and A. Traganitis are with the Institute of Computer Science, Foundation for Research and Technology - Hellas (FORTH). N. Pappas is with the Department of Science and Technology, Link\"{o}ping University, Campus Norrk\"{o}ping, Sweden.
	E-mails: \{gstam, alfrag, tragani\}@ics.forth.gr, nikolaos.pappas@liu.se.\\This research has been co-financed by the European Region Development Fund of the European Union and Greek national funds through the Operational Program Competitiveness, Entrepreneurship and Innovation, under the  call RESEARCH–CREATE–INNOVATE (project code: T1EDK-00070). The work of N. Pappas was supported in part by the Swedish Research Council (VR), ELLIIT, and CENIIT.}}

\maketitle
\vspace{-50pt}
\begin{abstract}
With its growing number of deployed devices and applications, the Internet of Things (IoT) raises significant challenges for network maintenance procedures. 
In this work we address a problem of active fault detection in an IoT scenario, whereby a monitor can probe a remote device in order to acquire fresh information and facilitate fault detection.
However, probing could have a significant impact on the system's energy and communication resources.
To this end, we utilize Age of Information as a measure of the freshness of information at the monitor and adopt a semantics-aware communication approach between the monitor and the remote device. 
In semantics-aware communications, the processes of generating and transmitting information are treated jointly to consider the importance of information and the purpose of communication.
We formulate the problem as a Partially Observable Markov Decision Process and show analytically that the optimal policy is of a threshold type. 
Finally, we use a computationally efficient stochastic approximation algorithm to approximate the optimal policy and present numerical results that exhibit the advantage of our approach compared to a conventional delay-based probing policy.
\end{abstract}

\section{Introduction}
\label{sec:introduction}
The emergence of massive IoT ecosystems poses new challenges for their maintenance procedures. 
IoT networks are characterized by software, hardware, and communication protocols' diversity. 
Furthermore, they are typically comprised of a large number of devices that are often deployed in remote and harsh environments.
In this context, the development of autonomous fault detection procedures is necessary to safely and efficiently operate an IoT network.
The majority of fault detection algorithms that have been proposed in the past~\cite{muhammed2017analysis, zhang2018survey}, assume that the system is passively monitored and utilize statistical or machine learning techniques to infer the actual health status of its subsystems. 
However, a major drawback with passive monitoring is that faults can pass undetected if the faulty and the nominal operation overlap due to measurement and process uncertainties or in cases where control actions mask the influence of faults~\cite{campbell2015auxiliary}.
To address this problem we make use of an \emph{active} fault detection scheme that utilizes probes to affect the system's response and thus to increase the probability of detecting certain faults.

With active fault detection special care must be taken so that the extra network traffic due to probing is not detrimental to the system's performance.
This requirement can become a significant challenge if active fault detection is to be deployed in IoT networks with a large number of remote devices.
Blindly generating and transmitting probes could increase network congestion and prohibit other applications from satisfying their possibly strict real-time constraints.
To this end, we adopt a \emph{semantics}-aware~\cite{pappas_kountouris_2021semantics} approach to active fault detection.
Within the context of \emph{semantics}-aware communications the generation and transfer of information across a network are considered jointly in order to take into account the goal or purpose of the communication.
What is more, the \emph{importance/significance} of a communication event, i.e., the event of generation and transmission of information, constitutes the decisive criterion of whether it should take place or not.
The definition of the importance of a communication event is application-specific, thus, in the context of active fault detection, we define it to be a function of the freshness of information that has been received from the remote device and of the operational status of the communication network and the remote device.
To put it simply, the importance of a probe increases when the information sent by the remote device has become stale and the probing entity is not confident that the operational status of the communication network and the remote device is good.
Numerical results indicate that, in contrast to the classical communication paradigm where information generation and its transmission are treated separately, the semantics-aware approach offers significant advantages.

More specifically, in this work, we consider a basic active fault detection scenario for a discrete-time dynamic system that is comprised of a sensor and a monitor. At the beginning of each time slot, the sensor probabilistically generates and transmits status updates to the monitor over an unreliable link while the monitor decides whether or not to probe the sensor for a mandatory transmission of a fresh status update through a separate unreliable link.
By the end of each time slot, the monitor may or may not receive a status update either because none was generated at the sensor or due to intermittent faults at the sensor and the wireless links.
To detect intermittent faults, the monitor maintains a belief vector, i.e., a probability distribution, over the operational status (healthy or faulty) of the system and a measure of its confidence in this belief vector that is expressed by the entropy of the belief vector.
Probing, successfully or unsuccessfully, increases the confidence of the monitor in its belief state, however, it also induces a cost for the monitor that measures the negative impact of probing on the system's energy and communication resources.
Our objective is to find a policy that decides at each time slot whether or not a probe should be sent to the sensor so that it optimally balances the probing cost with the need for fresh information at the monitor. 

Our approach to solving this problem is to formulate it as a Partially Observable Markov Decision Process (POMDP) and derive the necessary conditions for probing to result in a reduction of the belief state's vector entropy.
To the best of our knowledge, this is the first work with this approach. 
Our analysis indicates that there exist probing cost values such that the optimal policy is of a \emph{threshold} type.
In addition, we propose a stochastic approximation algorithm that can compute such a policy and, subsequently, evaluate the derived policy numerically.
 
\subsection{Related work}
Fault detection methods can be categorized as passive, reactive, proactive, and active.
Passive fault detection methods collect information from the data packets that the wireless sensors exchange as part of their normal operation whereas in reactive and proactive fault detection methods the wireless sensors collect information related to their operational status and subsequently transmit it to the monitor. 
Finally, in active fault detection methods the monitor probes the wireless sensors for information specific to the fault detection process.
In Wireless Sensor Networks (WSNs) the fault detection algorithms being cited in recent works and surveys~\cite{zhang2018survey, effah2018survey, mehmood2018survey, 9484556} fall in the passive, reactive, and proactive categories, with the majority of them being passive fault detection algorithms.
Active fault detection methods for WSNs have received limited attention~\cite{mehmood2018survey} compared to the other three categories. 
In~\cite{khan2008dustminer} and~\cite{yang2007clairvoyant} the authors adopted an active approach to fault detection in WSNs.
However, both of these tools were meant for pre-deployment testing of WSNs software rather than a health status monitoring mechanism. 

Unlike these works, we propose an active fault detection method for continuously monitoring the health status of sensors. 
We believe that autonomous active fault detection methods can successfully complement the passive ones by addressing their limitations. 
More specifically, passive fault detection methods often fail to detect faults because the faulty and the nominal operation overlap due to measurement and process uncertainties. 
What is more, network control mechanisms specifically designed to increase the robustness of the IoT network, e.g., by delegating the job of a senor to neighboring or redundant nodes, often compensate for the performance degradation due to intermittent faults and thus mask their influence rendering them undetectable~\cite{campbell2015auxiliary}.
Acknowledging the fact that the network overhead due to active fault detection can be prohibitive, we adopted the semantics-aware communication paradigm ~\cite{pappas_kountouris_2021semantics, popovski2020semantic, Strinati20216g, GCW2019Alarms, ICAS21, 9663101, 9513758, 9679803} which has exhibited its ability to eliminate the transmission of redundant and uninformative data and thus minimize the induced overhead.

Active fault detection methods have also been studied in the context of wired networks~\cite{jeswani2015adaptive, tayal2018congestion, tayal2020traffic, hu2014need, quan2012detecting, jeswani2012adaptive}.  
However, the operational conditions of wired networks differ considerably from these of WSNs in terms of protocols, energy, bandwidth, and transmission errors so that techniques proposed for wired networks cannot be applied in WSNs.

\section{System Model}
\label{sec:systemModel}
\begin{figure}[h]
	\centering
	\includegraphics[scale=0.9]{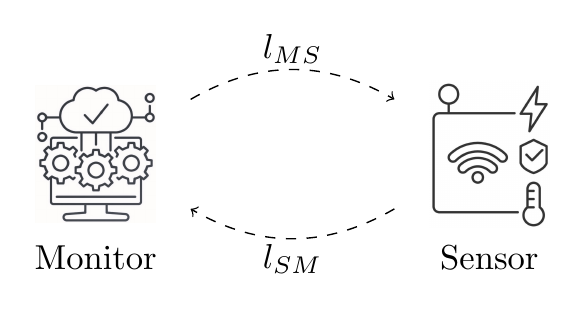}
	\caption{Basic IoT setup.}
	\label{fig:systemDiagram}	
	\vspace{-25pt}
\end{figure}

We consider the system presented in Figure~\ref{fig:systemDiagram}. 
It is comprised of a sensor that transmits status updates to a monitoring device over the wireless link labeled $SM$.
The monitoring device, besides receiving the status updates from the sensor, is able to probe the sensor for a fresh status update over the wireless link labeled $MS$.
Transmissions over the links $MS$ and $SM$ are subject to failure. 
Failures are independent between the two links.
We assume that time is slotted and indexed by $t \in \mathbb{Z}^+$.
\vspace{-25pt}

The state of the sensor is modeled as an independent two state time-homogeneous Markov process. 
Let $F_t^S \in \{0,1\}$ be the state of the sensor's Markov process at the beginning of the $t$-th time slot. 
When $F_t^S$ has a value of 0/1, the $i$-th sensor's operational status is healthy/faulty.
We assume that the sensor will remain in the same state for the duration of a time-slot and, afterwards, it will make a probabilistic transition to another state as dictated by the state transition probability matrix $P^S$.
Furthermore, at the beginning of each time slot the sensor will generate a status update with probability $P_g$, when in a healthy state, while it will not generate a status update when in a faulty state. 
In this work we assume that $P_g < 1$, otherwise the probing system is redundant. 
In the case of a status update generation the sensor will transmit it over the $SM$ link. 
At the end of the time slot the status update is discarded independently from the outcome of the transmission. 

Similarly, we model the health status of the wireless links as two independent two-state time-homogeneous Markov processes.
Let $F_t^{MS}, F_t^{SM} \in \{0,1\}$ denote the state of the independent Markov processes for the $MS$ and $SM$ links respectively, at the beginning of the $t$-th time slot. 
When $F_t^{MS}$ and $F_t^{SM}$ take a value of 0/1 the operation of the wireless links is healthy/faulty. 
We assume that the wireless links will remain in the same state for the duration of a single time-slot and, subsequently, they will make a transition to another state as dictated by the transition probability matrices $P^{MS}$ and $P^{SM}$ respectively. 
When in a healthy state, the wireless link will forward successfully a status update to the monitor with probability $1$, whereas a faulty wireless link will always fail to deliver the status update. 

In this work, we consider the problem of an autonomous agent that optimally probes the sensor to maximize the aggregate Value of Information (VoI) over a finite horizon. 
The VoI metric depends on how confident the agent is about the system's health status as well as on the freshness of the status updates it receives. 
Since the agent doesn't have access to the true health status of the system it has to 
maintain a belief state $P_t$, i.e., a probability distribution over all possible states of the system at time $t$, based on the observations it makes, i.e., the arrival of status updates.
Using $P_t$, the agent derives $P_t^h$, which represents its belief about the health status of the subsystem of Figure~\ref{fig:systemDiagram} that is comprised of the sensor and the $l_{SM}$ link, i.e., the subsystem  that is responsible for the generation and transmission of the status updates. 
In the following sections we will refer to this subsystem as the Generation-Transmission (GT) subsystem.
$P_t^h$ is a probability distribution over two states that correspond to a healthy and a faulty GT-subsystem respectively. 
We represent the confidence level of the agent, regarding the health status of the GT-subsystem, with the information entropy of $P_t^h$, i.e., $H(P_t^h)$, which, for simplicity, we denote with $H_t$. 
The derivation of $P_t^h$ from $P_t$ and the formula for the computation of $H_t$ are presented in Section~\ref{sec:belief_state} right after the analytical definition of the \emph{belief state}.
In the following sections we will refer to $H_t$ as the health status entropy of the belief state $P_t$. 
A low entropy value means that the agent is confident about the health status of the GT-subsystem while a high entropy value means that the agent is less confident about it.  

Furthermore, to characterize the freshness of the status updates received at the monitor we utilize the Age of Information (AoI) metric that has received significant attention in the research community~\cite{NOW-AoI, sunmodiano2019age, yates2020age, 9598864, 9068225, 9565910, 8778671, 9374461, 9085402}. 
AoI was defined in \cite{kaul2012real} as the time that has elapsed since the generation of the last status update that has been successfully decoded by the destination, i.e., $\Delta(t) = t - U(t)$,
where $U(t)$ is the time-stamp of the last packet received at the destination at time $t$.
We use $\Delta_t,\ t = 0, 1, \dots, N$, to denote the AoI of the sensor at time $t$.
However, as the time horizon of the optimal probing problem increases $\Delta_t$ could assume values that would be disproportionately larger than $H_t$.   
To alleviate this problem we will use a normalized value of the AoI which we define as, $ \bar{\Delta}_t= \frac{\Delta_t}{N}, $
where $N$ is the length of the finite horizon measured in time-slots. 
Finally, VoI is defined as,
\begin{equation} \label{eq:definition_VoI}
	\setstretch{1.0}\
	V_t = \lambda_1 H_t + \lambda_2 \bar{\Delta}_t,
\end{equation}
where $\lambda_1$ and $\lambda_2$ are weights that determine the relative importance of each component of the metric.

At the beginning of each time-slot the agent will consider its belief about the system's state and the VoI of the system, and it will decide whether to probe the sensor for a fresh status update or not. 
If the agent decides to send a probe it will pay a cost of $c$ units.
Following a successful reception of a probe through the $MS$ link, the sensor will generate a fresh status update at the next time slot with probability 1, if it is in a healthy state; and with probability $0$ if it is in a faulty state.

\section{Problem Formulation}\label{sec:problemFormulation}
In this section we formulate the decision problem presented above as a Partially Observable Markov Decision Process (POMDP) denoted with $\mathcal{P}$. A POMDP model with a finite horizon $N$ is a 7-tuple $(\mathcal{S}, \mathcal{A}, Z, P, r, g, g_N)$, where $\mathcal{S}$ is a set of states, $\mathcal{A}$ is a set of actions, $Z$ is a set of possible observations, $P$ is a probability matrix representing the conditional transition probabilities between states, $r$ represents the observation probabilities, $g$ is the transition reward function and $g_N$ is the terminal cost incurred at the last decision stage. 
In the remaining part of this section we present the individual elements of $\mathcal{P}$ and formulate the corresponding dynamic program based on a belief state formulation~\cite{B17}. 

\subsubsection*{\textbf{State Space ($\mathcal{S}$)}}
At the beginning of the $t$-th time-slot the health state of the system is represented by the column vector, $ s_t = [F_t^{MS}, F_t^S, F_t^{SM}]^T $
where, as described in Section~\ref{sec:systemModel},  $F_t^i \in\{0, 1\}, i \in \{MS, S, SM\}$ indicate the health status of the wireless links ($MS$, $SM$) and of the sensor $S$, while $T$ is the transpose operator. 
By the definition of the system's state we have eight discrete states which are indexed by $i = 0, 1, \dots,  7$.
The true state of the system is unknown to the agent at time $t$.

\subsubsection*{\textbf{Actions ($\mathcal{A}$)}}
The set of actions available to the agent is denoted with $\mathcal{A} = \{0, 1\}$, where $0$ represents the no-probe action and $1$ indicates the probe action. 
The result of the probe action, given that the probe is successfully received through the $MS$ link, is that the sensor will generate a fresh status update at the next time slot w.p. 1, if it is in a healthy state, and w.p. $0$ if it is in a faulty state, respectively.
Both actions are available in all system states. 
Finally, we denote the action taken by the agent at the beginning of the $t$-th time slot with $a_t \in \mathcal{A}$.
The action taken by the agent does not directly affect the state of the system, nevertheless it affects the observation made by the agent. 

\subsubsection*{\textbf{Random variables}}
\label{sec:randomVariables}
The state of the system presented in Fig.~\ref{fig:systemDiagram} will change stochastically at the beginning of each time slot.  
The transition to the new state is governed by the transition probability matrices $P^{MS}$, $P^{S}$ and $P^{SM}$, that were presented in Section~\ref{sec:systemModel}, and the state of the system during the previous time slot. 

As mentioned above, the agent has no knowledge of the system's actual state and is limited to observing the arrival of status updates. 
The observations are stochastic in nature and are determined by the action taken by the agent, the state of the system and the following random variables.
The random variable $W_t^g \in \{0, 1\}$ represents the random event of a status update generation at the $t$-th time slot. 
If a status update is generated by the sensor then $W_t^g$ takes the value $1$ and if the sensor does not generate a status update then $W_t^g$ takes the value $0$.
We have the following conditional distribution for $W_t^g$
\begin{equation} 
	\setstretch{1.0}\
	P[W^g = 0 | F^{S}, a] = 
	\begin{cases}
		1 - P_g, & \text{if } a = 0 \text{ and }  F^{S} = 0, \\
		1, & \text{if } a = 0 \text{ and }  F^{S} = 1 \text{, or } \text{if } a = 1  \text{ and } F^{S} = 1  \\
		0, & \text{if } a = 1 \text{ and }  F^{S} = 0, \\
	\end{cases}
\end{equation}
and $P[W^g = 1 | F^{S}, a] = 1 - P[W^g = 0 | F^{S}, a]$,
where we omitted the time index since the distribution is assumed to remain constant over time.

The random variable $W_t^{MS} \in \{0,1\}$ represents the random event of a successful transmission over the $MS$ link during the $t$-th time slot.
A value of $0$ indicates an unsuccessful transmission over the link and a value of $1$ indicates a successful transmission.
The conditional probability distribution for $W_t^{MS}$ is given by,
\begin{equation}
	\setstretch{1.0}\
	P[W^{MS} = 0 | F^{MS}, a] = 
	\begin{cases}
		1, & \text{if } a = 0 \text{ or } (a = 1  \text{ and } F^{MS} = 1) \\
		0, & \text{if } a = 1 \text{ and }  F^{MS} = 0, \\
	\end{cases}
\end{equation}
and $	P[W^{MS} = 1 | F^{MS}, a] = 1 - P[W^{MS} = 0 | F^{MS}, a]$,
where again we omitted the time index $t$.
Finally, the random variable $W_t^{SM} \in \{0,1\}$, represents the random event of a successful transmission over the $SM$ link during the $t$-th time slot. A value of $0$ indicates an unsuccessful transmission over the link and a value of $1$ indicates a successful transmission.
The conditional probability distribution for $W^{SM}$ is given by,
\begin{equation}
	\setstretch{1.0}\	P[W^{SM} = 0 | W^g, F^{SM}] = 
	\begin{cases}
		1, & \text{if } W^g = 0 \text{ or } (W^g = 1  \text{ and } F^{SM} = 1),\\
		0, & \text{if } W^g = 1 \text{ and }  F^{SM} = 0, \\
	\end{cases}
\end{equation}
and $P[W^{SM} = 1 | W^g, F^{SM}] = 1-P[W^{SM} = 0 | W^g, F^{SM}]$.

\subsubsection*{\textbf{Transition probabilities ($P$)}}\label{sec:systemDynamics}
Let $m$ be an index over the set of the three subsystems presented in Fig.~\ref{fig:systemDiagram}, i.e., $m \in \{MS, S, SM\}$, then the transition probability matrices $P^{MS}$, $P^{S}$ and $P^{SM}$ can be defined as follows, $ P^{m} = \left[ \begin{array}{cc}
	p_{00}^{m} & p_{01}^{m} \\
	p_{10}^{m} & p_{11}^{m} \end{array} \right]$,
where $p_{00}^m$ represents the probability to make a transition from a healthy state (0) to a healthy state (0) for subsystem $m$. 
Transition probabilities $p_{01}^{m}$, $p_{10}^{m}$, and $p_{11}^{m}$ are defined in a similar way.
Furthermore, we introduce the shorthand notation $s = [s_0, s_1, s_2]$ and $s'= [s_0', s_1', s_2']$ for states $s_t = [F_t^{MS}, F_t^S, F_t^{SM}]^T$ and $s_{t+1}$ respectively so that the conditional probability distribution of state $s'$ given the current state $s$ can be expressed as, $ P[s_{t+1} = s'|s_t = s] = p_{s_0 s_0'}^{MS} \cdot p_{s_1 s_1'}^{S} \cdot p_{s_2 s_2'}^{SM} $.

\subsubsection*{\textbf{Observations ($Z$)}}
At the beginning of each time slot the agent observes whether a status update arrived or not. 
Let $z_t \in \{0, 1\}$ denote the observation made at the $t$-th time slot, with $0$ representing the event that no status update was received and $1$ representing the event that a status update was received.
We define $r_{s}(a,z)$ as the probability to make observation $z$ at the $t$-th time slot, i.e.,  $z_t = z$, given that the system is in state $s$, i.e., $s_t = s$, and the preceding action was $a$, i.e., $a_{t-1} = a$. Thus we have, $r_{s}(a,z)  =  P[z_t = z | s_t = s, a_{t-1} = a]$.
The expressions of $r_{s}(a,z)$ for all combinations of states and actions are presented in Expressions~(\ref{eqn:r_s_0_0}),~(\ref{eqn:r_s_0_1}),~(\ref{eqn:r_s_1_0}), and~(\ref{eqn:r_s_1_1}).
\begin{figure*}[!t]
	\small
	\begin{align}
		\label{eqn:r_s_0_0}
		r_s(0, 0) &=  P[W_t^g=0|F_t^S, 0] + P[W_t^g = 1 | F_t^S, 0] P[W_t^{SM} = 0 | F_t^{SM}, 0] \\
		\label{eqn:r_s_0_1}
		r_s(0, 1) &= P[W_t^g = 1| F_t^s, 0]  P[W_t^{SM}| F_t^{SM}, 0] \\
	\label{eqn:r_s_1_0}
	r_s(1,0) &=  P[W_t^{MS}=0| F_t^{MS}, 1] + P[W_t^{MS} = 0 | F_t^{MS}, 1] P[W_t^g=1|F_t^s, 1] P[W_t^{SM}=0| F_t^{SM}, 1] + \\ \nonumber
	& P[W_t^{MS}=1 | F_t^{MS}, 1] P[W_t^g = 0| F_t^S, 1] + 
	P[W_t^{MS} = 1| F_t^{MS}, 1] P[W_t^g=1|F_t^s, 1]  P[W_t^{SM} = 0| F_t^{SM}, 1] \\
	\label{eqn:r_s_1_1}
	r_s(1,1) &=  P[W_t^{MS} = 1| F_t^{MS}, 1]  P[W_t^g=1 | F_t^S, 1] P[W_t^{SM}=1 | F_t^{SM}, 1] +\\ \nonumber
	& P[W_t^{MS}=0|F_t^{MS}, 1]  P[W_t^g = 1 | F_t^S, 1]  P[W_t^{SM} = 1| F_t^{SM}, 1]
	\end{align}
	\hrulefill
	\vspace{-20pt}
\end{figure*}
By utilizing the conditional probability distributions presented in Section~\ref{sec:randomVariables} we derived the observation probabilities for all possible combinations of states and controls and present them in Table~\ref{tbl:observation_probabilities}.
\begin{table}
	\caption{Observation probabilities $r_{s}(a,z)$ as a function of the health status of the sensor ($F^{S}$), of the $MS$ link ($F^{MS}$), of the $SM$ link ($F^{SM}$), the action ($a_{t-1}$) and the observation $z_t$.}
	\centering
	\label{tbl:observation_probabilities}
	\small
	\begin{tabular}{|c c c c|c|c|c|c|c|}
		\hline
		\multicolumn{1}{|c}{} & \multicolumn{3}{c}{}  & \multicolumn{2}{|c|}{$a_{t-1} = 0$} & \multicolumn{2}{|c|}{$a_{t-1} = 1$}  \\ \cline{5-8}
		$i$ & $F_t^{MS}$ & $F_t^{S}$ & $F_t^{SM}$ & $z_t = 0$ & $z_t = 1$ & $z_t = 0$ & $z_t = 1$\\ 
		\hline
		\hline 
		0 & 0 &  0 &  0  &$1-P_g$  	&$P_g$  	&0  		&1 \\
		\hline 
		1 & 0 &  0 &  1  &1  			&0  	  	&1  		&0 \\
		\hline 	
		2 & 0 &  1 &  0  &$1$  	&$0$  	&$1$  &$0$ \\
		\hline  
		3 & 0 &  1 &  1  &1  			&0  	  	&1  		&0 \\
		\hline 
		4 & 1 &  0 &  0  &$1-P_g$  	&$P_g$ 	&$1-P_g$  &$P_g$ \\
		\hline 
		5 & 1 &  0 &  1  &1  			&0  	  	&1  		&0 \\
		\hline 
		6 & 1 &  1 &  0  &$1$  	&$0$  	&$1$  &$0$ \\
		\hline 
		7 & 1 &  1 &  1  &1  			&0  		&1  		&0 \\
		\hline
	\end{tabular}
\vspace{-30pt}
\end{table}

The evolution of the AoI value over time depends on the observation made by the agent and,
\begin{equation}
	\setstretch{1.0}
	\Delta_{t} = 
	\begin{cases}
		1, &	\text{if } z_t = 1 \\ 
		\min\{N, \Delta_t + 1\}, & \text{if } z_t = 0
	\end{cases}
\end{equation}
where $N$ is the finite time horizon of the optimization problem.

\subsubsection*{\textbf{Transition cost function ($g$, $g_N$)}}
At the end of each time slot, the agent is charged with a cost that depends on the VoI and the action taken by the agent as follows, $g_t = c\cdot \mathds{1}_{\{a_t = 1\}} + V_t$, 
where, $\mathds{1}_{\{a_t = 1\}}$ is the indicator function which takes a value of $1$ when the probe action was taken by the agent and a value of zero otherwise, and $V_t$ is computed using Equation~(\ref{eq:definition_VoI}).
Parameter $c$ is a cost value associated with probing and quantifies the consumption of system resources for the generation and transmission of a probe.
\emph{The use VoI as a cost metric is justified by the fact that it expresses how much the agent is in need for a fresh status update from the sensor}.

\subsubsection*{\textbf{Total cost function}}
In a POMDP the agent doesn't have access to the current state of the system, thus, to optimally select actions it must utilize all previous observations and actions up to time $t$~\cite[Chapter 4]{B17}.
Let $h_t = [z_0, z_1, \dots, z_t, a_0, a_1, \dots, a_{t-1}]$ be the \emph{history} of all previous observations and actions, with $h_0 = \{z_0\}$. 
Furthermore, let $\mathcal{H}$ be the set of all possible histories for the system at hand.
The agent must find a policy $\pi^*$ that maps each history in $\mathcal{H}$ to a probability distribution over actions, i.e., $\pi:~\mathcal{H} \rightarrow P(\mathcal{A})$, so that the expected value of the total cost accumulated over a horizon of $N$ time slots is minimized. 
Let $\Pi$ be the set of all feasible policies for the system at hand, then, assuming that the agent's policy is $\pi \in \Pi$ and has an initial history $h_0$ the expected value of the total cost accumulated over a horizon of $N$ time slots is,
\begin{equation}\label{eq:ValueFunction}
	J_{0}(h_0) = \mathop {\mathbb{E}}_{W_0, \cdots, W_{N}} \Big[ \sum_{t=0}^{N} g_t \big\vert h_0, \pi \Big],
\end{equation}
where expectation $\mathbb{E}\{\cdot\}$ is taken with respect to the joint distribution of the random variables in $W_t=[W_t^{MS}, W_t^{g}, W_t^{SM}]^T$ for $t= 0, 1,\dots$ and the given policy $\pi$. 
Our objective is to find the optimal policy $\pi^*$ which is defined as $ \pi^* = \arg\underset{\pi \in \Pi}{\min}\, J_{\pi, N}(h_0) $. 

For finite $N$ the optimal policy $\pi^*$ can be obtained by using the dynamic programming  algorithm.
However, the difficulty with this approach is that the dynamic programming algorithm is carried out over a state space of expanding dimension.
As new observations are made and new actions are taken the dimension of $h_t$ increases accordingly.
To overcome this difficulty $h_t$ can be replaced by a sufficient statistic, i.e., a quantity that summarizes all the essential content of $h_t$ that is necessary for control purposes.
In the POMDP literature a sufficient statistic that is often used is the belief state which is presented in the following section.

\subsubsection*{\textbf{Belief State}}\label{sec:belief_state}
At each time slot $t$ the agent maintains a belief state $P_t$, i.e., a probability distribution over all possible system states, $ P_t = [p_t^0, \dots, p_t^7]^T. $
Starting from an arbitrarily initialized belief state $P_0$ the agent updates its belief about the actual state of the system at the beginning of each time slot as follows,
\begin{equation} \label{eq:NextBeliefState}
	\setstretch{1.0}\
	p_{t+1}^j = \frac{\sum_{i=0}^{7} p_t^i \cdot p_{ij} \cdot r_j(a, z)}{\sum_{s=0}^7 \sum_{i=0}^7 p_t^i \cdot p_{is} \cdot r_s(a, z)},
\end{equation}
where $p_{ij} = P[s_{t+1} = j | s_t = i]$. In the literature $p_{ij}$ is usually a function of the action selected at time $t$, i.e., $p_{ij}(a_t)$, however, in our case the actions taken by the agent do not affect the system's state. 
In any case, the action taken by the agent affects the observation $z_{t+1}$ made by the agent and thus directly affects the evolution of the belief state over time.
As mentioned in Section~\ref{sec:introduction}, based on $P_t$ the agent forms the \emph{health status belief vector} $P_t^h$ that represents our belief regarding the health status of the sub-system comprised of the sensor and the $l_{SM}$ link.
We have, $P_t^h=[p_t^h, p_t^f]$, where $p_t^h$ and $p_t^f$ represent, respectively, the probabilities for the sub-system to be in a healthy or faulty state. 
We define $p_t^h = p_t^0 + p_t^4$, since states with index $0$ and $4$ in Table~\ref{tbl:observation_probabilities} are the only states where both the sensor and the $l_{SM}$ link are in a healthy state. 
Correspondingly, we define $p_t^f = \sum_{i \neq 0,4} p_t^i, \, i=0, \cdots, 7$. 
It holds that $P_t^h$ is a probability distribution since $p_t^h$ and $p_t^f$ are computed over complementary subsets of the system's state space and $P_t$ is a probability distribution.
Finally, the health status entropy is computed as $H_t = -[ p_t^h \cdot log_2(p_t^h) + p_t^f \cdot log_2(p_t^f)]$.

For the agent to have all the information necessary to proceed with the decision process it must also keep the value of the AoI as part of its state, thus we augment the belief state with the value of AoI and define the following representation of the current state, i.e., $ x_t = [P_t, \bar{\Delta}_t] $,
and define $X$ to be the set of all states.
\subsubsection*{\textbf{Dynamic program of $\mathcal{P}$}}
By utilizing the belief state formulation and for a finite horizon $N$ the optimal policy $\pi^*$ 
can be obtained by solving the following dynamic program, 
\begin{align}
	\label{eq:dynamicProgram}
	J_t(x_t) &= \min_{a_t \in \{0,1\}} \big[ g_t + \sum_z \sum_s \sum_i p_t^i\, p_{is} \, r_s(a_t,z)\, J_{t+1}(x_{t+1}) \big]
\end{align}
for all $x_t \in X$ and $t = 0, 1, \cdots N$, where $x_{t+1} = [P_{t+1}^{a, z}, \Delta_{t+1}^z]$, $z \in \{0,1\}$, $s, i \in \{0, 1, \dots, 7\}$ and the terminal cost is given by $J_{N}(x_{N}) = g_{N}$.
\emph{The formulation of~(\ref{eq:dynamicProgram}) differs from the typical dynamic program for the general case of a POMDP~\cite{krishnamurthy2016partially, B17}, due to the fact that the transition cost depends only on the observed values.} 

It is known that for~(\ref{eq:dynamicProgram}) there do exist optimal stationary policies~\cite{krishnamurthy2016partially, B17}, i.e., $\pi^* = \{\pi_0^*, \pi_1^*, ..., \pi_{N-1}^*\}$. However, since the state space $X$ is uncountable the recursion in~(\ref{eq:dynamicProgram}) does not translate into a practical algorithm.
Nevertheless, based on~(\ref{eq:dynamicProgram}) we can prove that the optimal policy has certain structural properties that can be utilized for its efficient computation.

\section{Analysis}
\label{sec:analysis}
In this section we present structural results for the optimal policy of the POMDP $\mathcal{P}$ defined in the previous section.  
In order to represent the belief state at the $(t+1)$-th time slot one has to consider the action that was taken at time $t$, i.e., $a_t$, and the observation made at $(t+1)$, i.e., $z_{t+1}$, thus, we use $P_{t+1}^{a,z}$ to represent the belief state at the $(t+1)$-th time slot, when $a_t = a$ and $z_{t+1} = z$. 
In this work we assume that POMDP $\mathcal{P}$ satisfies the following two assumptions.
\begin{assumption} \label{assmption:causalInEntropy}
Let $x_t = [P_t, \bar{\Delta}_t]$ and $x_t^{+} = [P_t^+, \bar{\Delta}_t]$ be states such that $H(P_t^{h,+}) \geq H(P_t^h)$ then $H(P_{t+1}^{h, +, a, z}) \geq H(P_{t+1}^{h, a, z})$, $a, z\in\{0,1\}$.
\end{assumption}
Assumption~\ref{assmption:causalInEntropy} states that the health status entropy $H(P_{t+1}^{h, a, z, +})$ of the belief state $P_{t+1}^{a, z, +}$, which results from belief state $P_t^+$ by taking action $a_t = a$ and subsequently observing $z_{t+1} = z$, will be larger than the health status entropy that would result if the system had started in belief state $P_t$, which has a lower health status entropy than $P_t^+$, given that the \emph{same} action and observation had been made in both cases. 
\begin{assumption} \label{assumption:NoObservationIndicatesFault}
	Let $I_S = \{0, 1, \dots, 7\}$ and $i \in I_S$ be the index of the system's state $s_t = [i_0, i_1, i_2]^T$ at time $t = 0, 1, \cdots N$, where $i_0 = F^{MS}$, $i_1 = F^S$, and $i_2 = F^{SM}$ (see Table~\ref{tbl:observation_probabilities}). Furthermore, let $p_{i_1 0}^{S}$, $p_{i_2 0}^{SM}$ be the probabilities for the sensor $S$ and the link $l_{SM}$ to make a transition from health status $i_1$ and $i_2$, respectively, to a healthy status (indicated by $0$) at $t+1$. We assume that for the POMDP $\mathcal{P}$ the following inequality is true,
	\begin{equation} \label{eq:lackImpliesFault}
	\sum_{i\in I_S} p_t^{i} \big[ p_{i_1 0}^{S} p_{i_2 0}^{SM} (2-P_g)  - 1 \big] \leq 0,\qquad t=0, 1, \dots, N.
	\end{equation}  
\end{assumption} 
Assumption~\ref{assumption:NoObservationIndicatesFault} expresses the necessary conditions and system's parametrization for the probing action to always result in a lower health status entropy compared to the no probe action.  
It may seem intuitive that probing reduces entropy, since it makes the generation of a status update from the sensor mandatory,  i.e., it reduces the uncertainty induced in the system due to the probabilistic generation of status updates from the sensor, however, one should also consider that probing introduces a new type of uncertainty in the system due to the transmission failures occurring in the $MS$ link. 
As an example consider the case where a probe was sent to the sensor yet no status update was received by the monitor. 
It is not certain whether this happened because the probe didn't actually reach the sensor, due to a faulty $MS$ link, or because the sensor, or the $SM$ link, or both were in a faulty state. 
Assumption~\ref{assumption:NoObservationIndicatesFault} expresses the effect of faults in the $MS$ link along with that of parameters $p_{i_1 0}^{S}$, $p_{i_2 0}^{SM}$ and $P_g$ on the health status entropy (for details see Appendices~\ref{apdx:ProbeDecrementsEntropy} and~\ref{apdx:xi_and_x2_leq_phi_s})
and it is used in the following lemma to prove that the probe action will always result in the same or reduced health status entropy compared to the no-probe action for a given observation $z$ at time ${t+1}$.
\begin{lemma} \label{lemma:ProbeDecrementsEntropy}
	Let $P_{t+1}^{0,z}$ and $P_{t+1}^{1 ,z}$ be the belief states of $\mathcal{P}$ at the $(t+1)$-th time slot when $a_t = 0$ and $1$, respectively, and let $P_{t+1}^{h, 0, z}$, $P_{t+1}^{h, 1,z}$ be their corresponding health status belief vectors, then, if Assumption~\ref{assumption:NoObservationIndicatesFault} is satisfied, it holds that, $H(P_{t+1}^{h, 0, z}) \geq H(P_{t+1}^{h, 1,z}),\quad z\in \{0,1\}. $
\end{lemma}
The proof is given in Appendix~\ref{apdx:ProbeDecrementsEntropy}.
Next, in Lemma~\ref{lemma:increasingInEntropy}, we show that the expected cost-to-go from decision stage $t$ up to $N$ is an increasing function of the health status entropy. 
\begin{lemma} \label{lemma:increasingInEntropy}
	Let $x_t^+ = [P_t^+, \bar{\Delta}_t]$ and $x_t^- = [P_t^-, \bar{\Delta}_t]$ be states such that $H(P_t^{h, +}) \geq H(P_t^{h,-})$ and $J_t(\cdot)$ be the dynamic program of $\mathcal{P}$ then for $t=1,\dots, N$, it holds that $J_t(P_t^+, \bar{\Delta}_t)  \geq J_t(P_t^-, \bar{\Delta}_t).$
\end{lemma}
\emph{Proof}: The proof of Lemma~\ref{lemma:increasingInEntropy} is given in Appendix~\ref{apdx:increasingInEntropy}.

In Lemma~\ref{lemma:increasingInAoI} we state a similar property for the expected cost-to-go when the value of AoI increases. We omit the proof of Lemma~\ref{lemma:increasingInAoI} since it is intuitive and its proof follows a similar line of arguments as in Lemma~\ref{lemma:increasingInEntropy}.
\begin{lemma} \label{lemma:increasingInAoI}
	Let $\bar{\Delta}_t^+$ and $\bar{\Delta}_t^-$ be normalized AoI values such that $\bar{\Delta}_t^+ \geq \bar{\Delta}_t^-$ and $J_t(\cdot)$ be the cost-to-go function in the dynamic program~(\ref{eq:dynamicProgram}) then for $t=0, 1,\dots, N-1$, it holds that $J_t(P_t, \bar{\Delta}_t^+)  \geq J_t(P_t, \bar{\Delta}_t^-)$.
\end{lemma}

In Lemma~\ref{lemma:linearJinEntropy} we prove properties of the cost-to-go function $J_t(\cdot)$ that are necessary to establish the structural properties of the optimal policy in Theorem~\ref{thm:monotonePolicy}
.\begin{lemma} \label{lemma:linearJinEntropy}
Let $J_t(x_t)$ be the value of the dynamic program of $\mathcal{P}$ at $x_t = [P_t, \bar{\Delta}_t]$ then $J_t(x_t)$ is piece-wise linear, increasing, and concave with respect to $H(P_t^h)$ and $\bar{\Delta}$ for $t= 1,\dots, N$.
\end{lemma} 
\emph{Proof}: The proof of Lemma~\ref{lemma:linearJinEntropy} is given in Appendix~\ref{apdx:linearJinEntropy}.

Finally, in Theorem~\ref{thm:monotonePolicy} we show that there exist configurations of POMDP $\mathcal{P}$ such that the optimal policy is threshold based. 
\begin{theorem} \label{thm:monotonePolicy}
	At each decision stage $t = 0, 1, \cdots, N-1$ there exists a positive probing cost $c$ such that the probing action is optimal for state $x_t^T = [P_t, \bar{\Delta}_t]$ and for all states $x_t^{+} = [P_t^{+}, \bar{\Delta}_t^+]$ with $H(P_t^{h,+}) \geq H(P_t^h)$ and $\bar{\Delta}_t^+ \geq \bar{\Delta}_t$.
\end{theorem}
\emph{Proof:} The proof of Theorem~\ref{thm:monotonePolicy} is given in Appendix~\ref{apdx:monotonePolicy}. 
\section{Optimal policy approximation}
According to Theorem~\ref{thm:monotonePolicy}, given a proper probing cost $c$, the optimal policy $\pi^* = \{\pi_0^*, \pi_1^*, \dots, \pi_{N-1}^*\}$ for the finite horizon POMDP $\mathcal{P}$ is of a threshold type. 
This means that  $\pi^* $  is comprised of different threshold values at \emph{each} decision stage $t = 0, 1, \dots, N$.
More specifically, let $\theta_t^{H,*}$ and $\theta_t^{\Delta,*}$ be the optimal threshold values for the health status entropy and the normalized AoI at stage $t$, then the optimal policy can be expressed as $\pi^* = \{[\theta_0^{H,*}, \theta_0^{\Delta,*}], [\theta_1^{H,*}, \theta_1^{\Delta,*}], \dots, [\theta_{N-1}^{H,*}, \theta_{N-1}^{\Delta,*}]\}$.
Computing $\theta_k^* = [\theta_k^{H,*}, \theta_k^{\Delta, *}]^T$ for $t =0, 1, \dots, N$ can be a computationally demanding task, especially if one considers large time horizons. 
To address this problem we approximate the optimal policy $\pi^*$ with a \emph{single} threshold and utilize a Policy Gradient algorithm, namely, the Simultaneous Perturbation Stochastic Approximation (SPSA) Algorithm~\cite{spall2005introduction} in order to find it.

The SPSA algorithm appears in Algorithm~\ref{alg:spsa} and operates by generating a sequence of threshold estimates, $\theta_k = [\theta_k^H, \theta_k^{\Delta}]^T$, $k = 1, 2, \dots, K$ that converges to a local minimum, i.e., an approximation of the best single threshold policy for POMDP $ \mathcal{P} $.
The SPSA algorithm picks a single random direction $\omega_k$ along which the derivative is evaluated at each step $k$, i.e., $ \omega_k^H $ and $ \omega_k^{\bar{\Delta}} $ are independently generated according to a Bernoulli distribution as presented in line~\ref{state:randomDirection} of Algorithm~\ref{alg:spsa}.
Subsequently, in line~\ref{state:thetaPlus} 
the algorithm generates threshold vectors $ \theta_k^+ $ and $ \theta_k^- $, which are bounded element-wise in the interval $[0, 1]$, i.e., $\mathbf{0}$ and $\mathbf{1}$ in line~\ref{state:thetaPlus} are column vectors whose elements are all zeros and ones respectively.
$ \theta_k^{\Delta} $ is also bounded in $[0, 1]$ since we assumed a normalized value for the AoI, and, this is also true for $ \theta_k^H $ since the maximum health status entropy occurs for $ P_t^h = [0.5, 0.5] $ which evaluates to $1$. 
In line~\ref{state:JhatPlus} 
the estimates $\hat{J}(\theta^+) $ and $\hat{J}(\theta^-) $ are computed by simulating  $M_s$ times the POMDP $\mathcal{P}$ under the corresponding single threshold policy. 
Finally, the gradient is estimated in line~\ref{state:gradientEstimate}, where $\oslash$ represents an element-by-element division, and $ \theta_k $ is updated in line~\ref{state:updateThreshold}.
Since the SPSA algorithm converges to local optima it is necessary to try several initial conditions $\theta_0$.
\begin{algorithm}[h!]
	\caption{Policy gradient algorithm for probing control}
	\label{alg:spsa}
	\begin{algorithmic}[1]
		\STATE Initialize threshold $\theta_0 = [\theta_0^H, \theta_0^{\Delta}]$ and $\gamma$, $A$, $\eta$, $\beta$, $\zeta$
		\FOR {$k$ = $1$ to $K$} 
			\STATE $ \gamma_k = \frac{\gamma}{(k+A)^{\beta}} $ and $ \eta_k = \frac{c}{k^{\zeta}} $ 
			\STATE \label{state:randomDirection} Randomly set $\omega_k^H$, $\omega_k^{\bar{\Delta}}$ to the equiprobable values $\{-1, 1\}$ and define $\omega_k = [\omega_k^H, \omega_k^{\bar{\Delta}}]^T$
			\STATE \label{state:thetaPlus} $ \theta_k^+ = \min \{\mathbf{1}, \max \{\mathbf{0}, \theta_{k-1} + \eta_k \cdot \omega_k \}\}  $ and $ \theta_k^- =  \min \{\mathbf{1}, \max \{\mathbf{0}, \theta_{k-1}  - \eta_k \cdot \omega_k \}\}$
			\STATE \label{state:JhatPlus} $ y_k^+ = \hat{J}(\theta_k^+) $, $ y_k^- = \hat{J}(\theta_k^-) $
			\STATE \label{state:gradientEstimate} $ \hat{e_k} = (y_k^+ - y_k^-) \oslash (2 \cdot c_k \cdot \omega_k)$
			\STATE \label{state:updateThreshold} $ \theta_k  = \theta_{k-1} - \gamma_k \hat{e} $
		\ENDFOR
	\end{algorithmic}
\end{algorithm}
\vspace{-30pt}
\section{Numerical Results}
In this section, we evaluate numerically the cost efficiency of the threshold probing policy we derived previously. 
Furthermore, we provide comparative results with an alternative probing policy that is often used in practice. 
The latter policy will probe the sensor whenever the time that has elapsed since the last arrival of a status update at the monitor exceeds a certain threshold.
We will refer to this policy as the \emph{delay} based policy whereas we will refer to the single threshold policy that approximates the optimal policy as the \emph{threshold} policy.
We also note here that for the system we consider in this work, the delay and AoI metrics coincide. 
This holds because the sensor does not buffer status updates, and the status update generation scheme is fixed.
A consequence of this observation is that the results we present in this section exhibit the comparative advantage in using VoI instead of AoI when deciding whether to probe or not, which shows that AoI itself cannot significantly capture the semantics of information except for timeliness.

\begin{figure}
	\centering
	\begin{minipage}{.49\textwidth}
		\centering
		\includegraphics[scale=0.55]{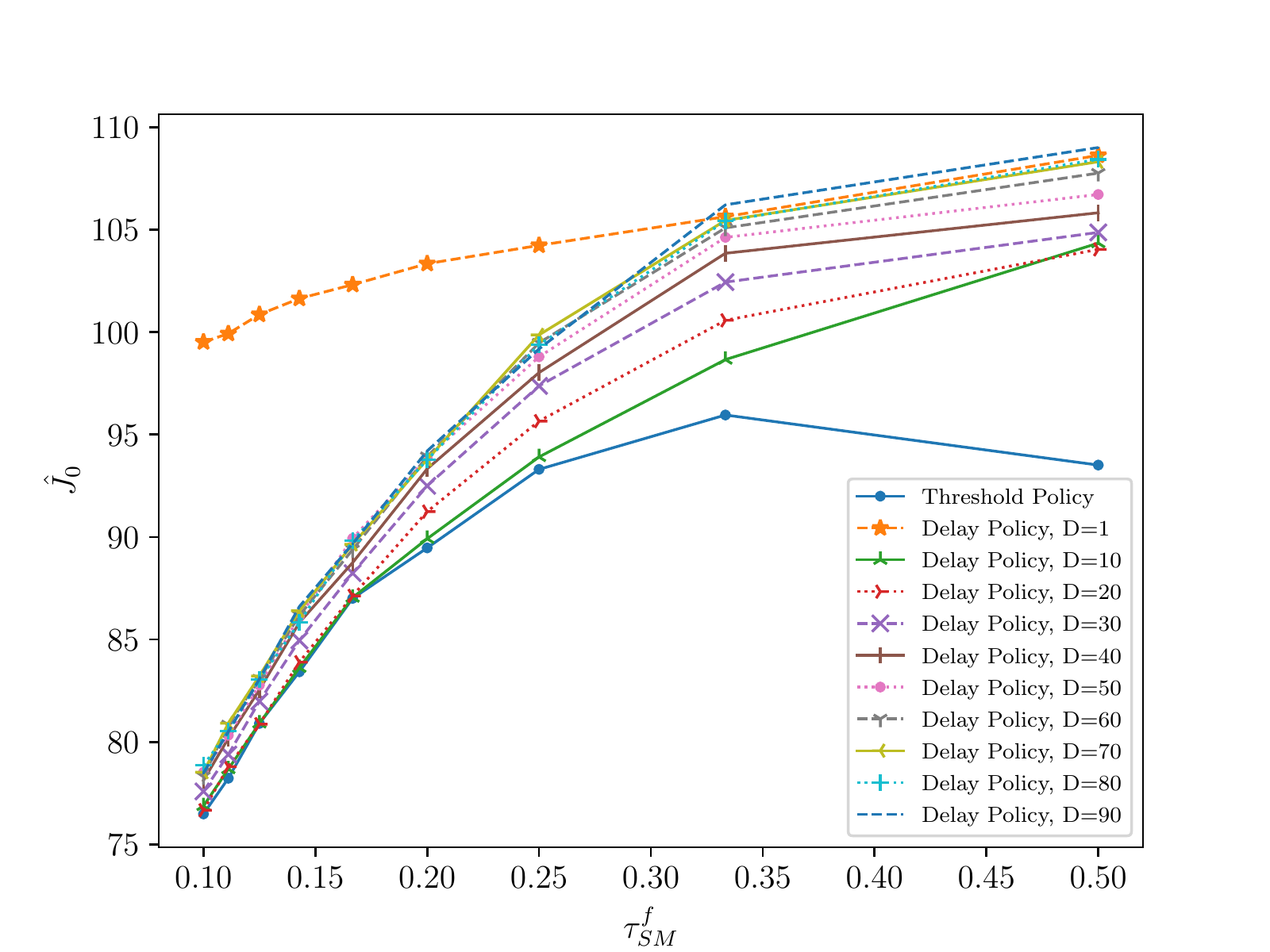}
		\caption{$\hat{J}_0(\cdot)$ vs $\tau_{SM}^f$ for a horizon of 100 time slots.}
		\label{fig:scenario_2}
	\end{minipage}%
	\hfill
	\begin{minipage}{.49\textwidth}
		\centering
		\includegraphics[scale=0.55]{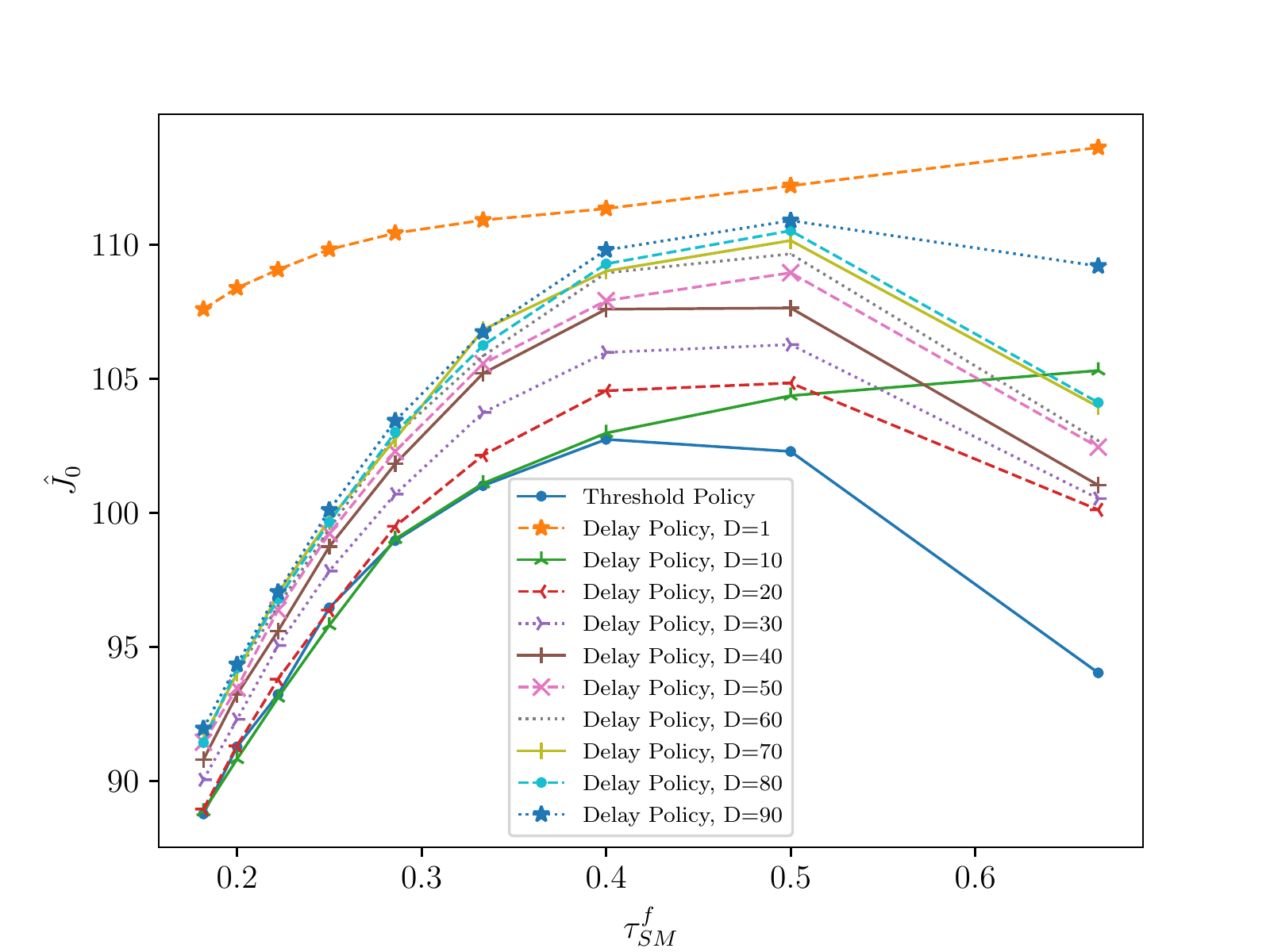}
		\caption{$J_0(\cdot)$ vs. $\tau_{SM}^f$ when we increase the probability for the sensor to enter a faulty state.}
		\label{fig:scenario_2_3}
	\end{minipage}
\vspace{-25pt}
\end{figure}

Furthermore, in order to gain insight into how the various system parameters affect the performance of the probing policies we formulated a \emph{basic} scenario and subsequently varied its parameters.
For this scenario, the system was configured as follows, $c = 1$, $\lambda_1 = 1$, $\lambda_2 = 1$, $P_g = 0.1$ and the transition probability kernels were set as,
$P^{MS} = \left[ \begin{array}{cc}					
	0.9 & 0.1 \\
	0.9 & 0.1
\end{array}\right], \quad
P^{S} = \begin{bmatrix} 
 	0.9 & 0.1 \\
 	0.9 & 0.1 
\end{bmatrix}, \text{ and } 
  P^{SM} = \begin{bmatrix} 
  	0.9 & 0.1 \\
  	1 - p_{SM}^{11} & p_{SM}^{11}
  \end{bmatrix}$,
where $p_{SM}^{11} = 0.1, 0.2, \dots, 0.9$. 
Furthermore, we set the parameters of the SPSA algorithm as follows, $\eta = 1$, $\gamma = 10^{-3}$,  $A = 1$, $\beta = 1$, and $\zeta = 1$. 
We derived the threshold policy by executing $K = 20$ iterations of the SPSA algorithm.
At each iteration $k = 1,2, \dots, K$ we calculated each of $y_k^+$ and $y_k^-$ as the sample average of $100$ Monte-Carlo simulations.
Each Monte-Carlo simulation had a time horizon of $N$ time slots and during that period the system was controlled by the single threshold policy defined by $\theta_k^+$, in the case of  $y_k^+$, and $\theta_k^-$, in the case of  $y_k^-$, as presented in Algorithm~\ref{alg:spsa}.
Subsequently, we used the threshold $\theta_K$ to evaluate the efficiency of the derived threshold policy.
More specifically, for all policies appearing in Figure~\ref{fig:scenario_2}, was calculated the average cost $\hat{J}_0$ as the sample average over $M = 2000$ Monte-Carlo simulations of the system while it was being controlled by the corresponding policy over a period of $N$ time slots, i.e., $ 	\hat{J}_{0} = \frac{1}{M}\sum_{m = 1}^{M} \sum_{t=0}^{N} g_t. $
Finally, for each Monte-Carlo simulation we set randomly the initial health status for the sensor and the $l_{MS}$ and $l_{SM}$ links.

In Figure~\ref{fig:scenario_2} we present the evolution of $\hat{J}_0$ with respect to the steady state probability of link $l_{SM}$ to be in a faulty state, $\tau_{SM}^f = \frac{p_{SM}^{01}}{1 - p_{SM}^{11} + p_{SM}^{01}}$.
We utilized $\tau_{SM}^f$ instead of $p_{MS}^{11}$ because it assumes a more intuitive interpretation, i.e., it expresses the expected time that link $l_{SM}$ would spend in the faulty state over a large time horizon.
In Figure~\ref{fig:scenario_2} we consider delay policies that will probe the sensor when a new status update arrival has been delayed for more than $D = 1, 10, 20, \dots, 90$ time slots.
The results presented in Figure~\ref{fig:scenario_2} indicate that the threshold based policy achieved a lower cost $\hat{J}_0$ compared to the delay based policies.
In order to provide insights into this result we have to point out the behavior of the two extreme delay policies, i.e., those with $D$ equal to $1$ and $90$.
The first one of probed the sensor most often than the other policies since $D$ assumed its smallest value and the sensor generated status updates with a small probability, i.e., $P_g = 0.1$.
On the other hand, the delay policy with $D = 90$ practically never probed the sensor since $90$ is close to the entire time horizon of $100$ time slots.

From Figure~\ref{fig:scenario_2} we see that when $\tau_{SM}^f$ was less than $0.20$ the delay policy with $D = 90$ and the threshold policy had similar cost efficiency.
This means that probing was rarely needed for that range of $\tau_{SM}^f$ values.
When $\tau_{SM}^f$ lied in the range between $0.20$ and $0.30$ the cost induced by the delay policy with $D = 90$ increased with a higher rate compared to all other policies. 
This indicates that probing became necessary in order to reduce cost $\hat{J}_0$.
This is evident also by the fact that the delay based policy with $D = 10$ performed closer to the threshold policy within this range of $\tau_{SM}^f$ values.
Finally, when $\tau_{SM}^f$ became larger than $0.30$ all delay policies saw an increment in their induced cost  $\hat{J}_0$ while the threshold based policy managed to reduce the value of $\hat{J}_0$. 
In this latter range of $\tau_{SM}^f$ values the periods that the $l_{SM}$ link was in a faulty state increased in duration due to the increasing $p_{SM}^{11}$ value.
As a consequence the time interval between status update arrivals  increased and all delay based policies engaged in probing the sensor when delay exceed their $D$ threshold. 
However, while the $l_{SM}$ link was in a faulty state, no status update could reach the monitor. 
As a result, and despite the persistent  probing of the sensor by the delay based policies, neither the health status entropy nor the normalized AoI could be reduced. 
This is particularly evident in the abrupt increase in cost for the delay policies with $D = 1$ and $ D = 10$ which persisted in probing the sensor with higher frequency and for longer periods due to their low values for $D$. 
On the other hand, the threshold based policy was able to avoid unnecessary probing by utilizing the health status entropy along with the normalized AoI, i.e, it would defer probing while it was confident that the system was in faulty state.

In Figure~\ref{fig:scenario_2_3} we present cost $\hat{J}_0$ for a wider range of values for $\tau_{SM}$. 
More specifically, we modified the basic scenario by increasing $p_{SM}^{01}$ from $0.1$ to $0.2$.
As a result, the $l_{SM}$ link entered more often its faulty state compared to the basic scenario and this provided for a wider range of $\tau_{SM}^f$ values.
All policies exhibited the same behavior as in the basic scenario for values of $\tau_{SM}^f$ up to $0.5$. 
However, when $\tau_{SM}^f$ got larger than $0.5$
we observed a reduction in the induced cost $\hat{J}_0$ for or all policies except for the delay based policies with $D = 1$ and $D = 10$.
The observed reduction in $\hat{J}_0$ was mainly due to the reduction in the cost induced by the health status entropy. 
More specifically, for large values  $\tau_{SM}^f$, i.e., for large values of  $p_{SM}^{11}$ and $p_{SM}^{01}$, the monitor was confident about the health status of the system, i.e., that the system is in a faulty state mainly due to the $l_{SM}$ link, and this resulted in a reduced health status entropy.
The delay policies with $D = 1$ and $D = 10$ increased the induced cost $\hat{J}_0$ by persistently probing the sensor while it was in a faulty state until they succeeded in the transmission of a status update. 
As expected, for large values of $\tau_{SM}^f$, this behavior resulted in a large number of unnecessary probes. 
Policies with a larger value of $D$ were also engaged into this type of probing albeit with a significantly less frequency.

\begin{figure}[!htb]
\centering
\begin{minipage}{.49\textwidth}
	\centering
	\includegraphics[scale=0.55]{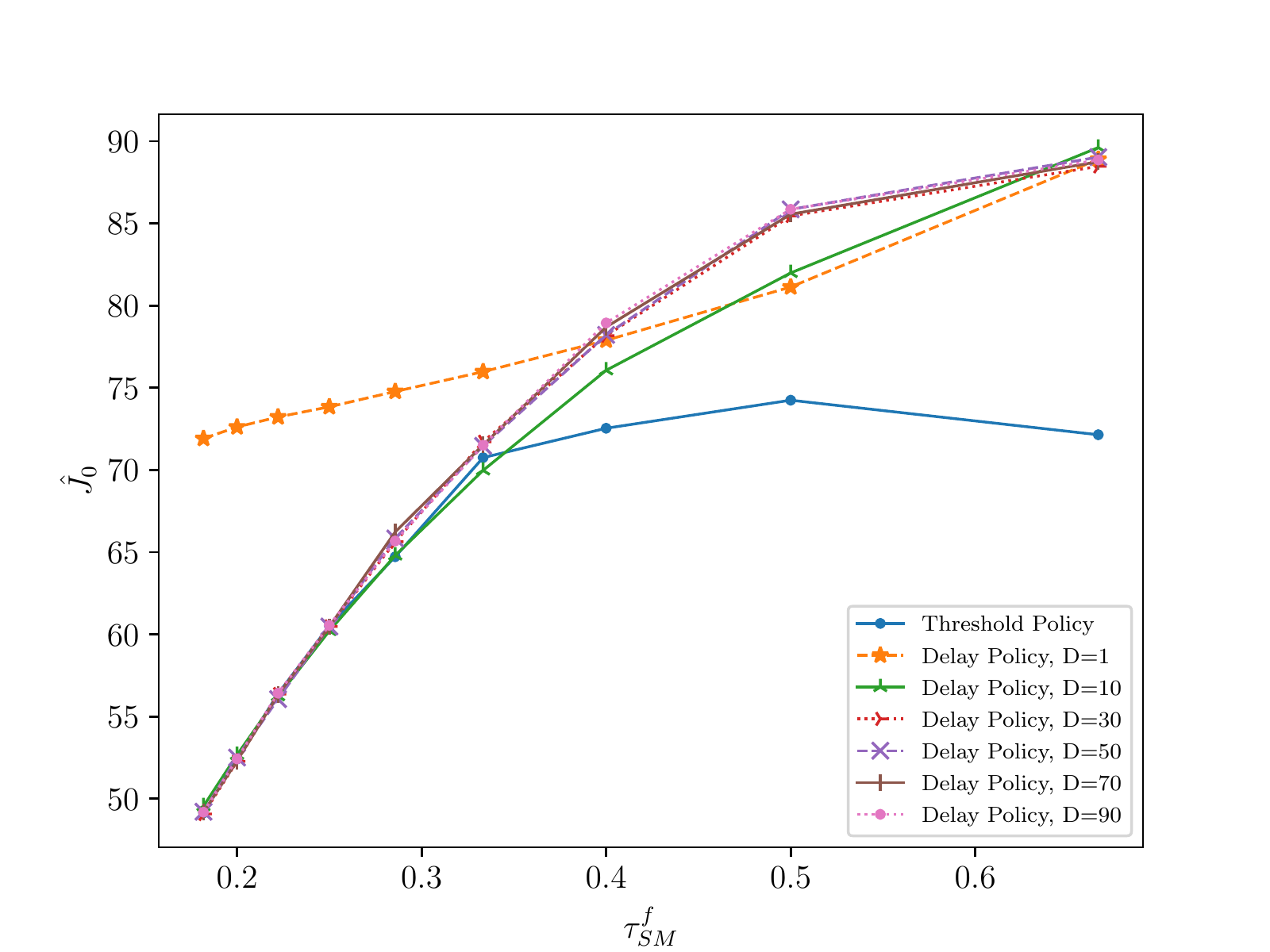}
	\caption{$J_0(\cdot)$ vs.  $\tau_{SM}^f$ when only the $l_{SM}$ link is subject to failures.}
	\label{fig:scenario_1_1}
\end{minipage}%
\hfill
\begin{minipage}{.49\textwidth}
	\centering
	\includegraphics[scale=0.55]{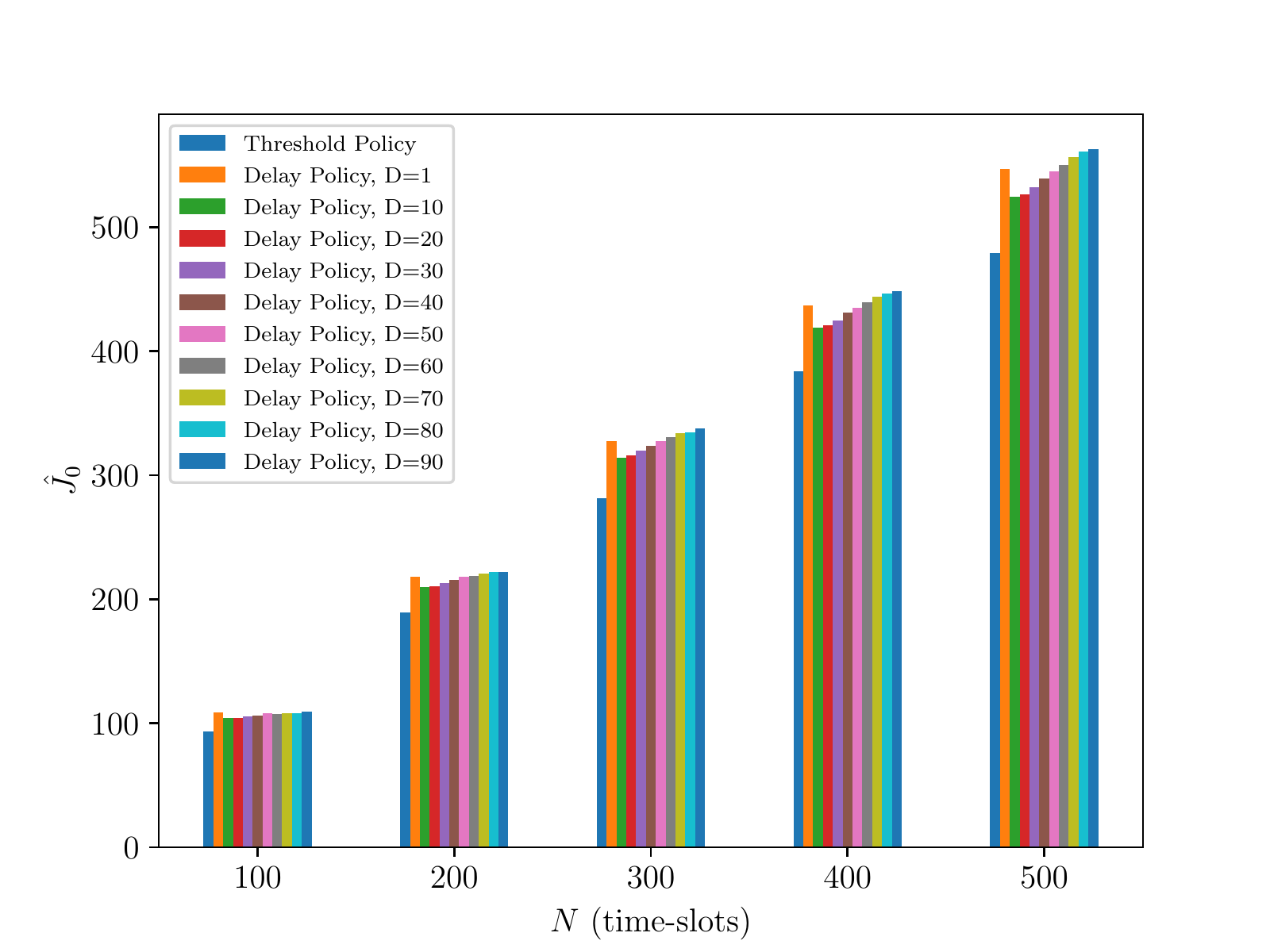}
	\caption{Sample average of cost $J_0(\cdot)$ vs the optimization horizon.}
	\label{fig:scenario_4}
\end{minipage}
\vspace{-25pt}
\end{figure}

In Figure~\ref{fig:scenario_1_1} we present the effect of the health status entropy on the cost $\hat{J}_0$ over all policies. 
We modified the basic scenario by setting,
$P^{MS} = \left[ \begin{array}{cc}					
		1 & 0 \\
		1 & 0
	\end{array}\right], 
P^{S} = \begin{bmatrix} 
		1 & 0 \\
		1 & 0 
	\end{bmatrix}$,
and by increasing $p_{SM}^{01}$ from $0.1$ to $0.2$ in order to get the same range of $\tau_{SM}^f$ values as in Figure~\ref{fig:scenario_2_3}.
By setting the matrices $P^{MS}$ and $P^{S}$ to the values presented above both the link $l_{MS}$ and the sensor $S$ would never enter a faulty state and, even if they were randomly initialized to a faulty state they would return to the healthy state with probability $1$ in the next time slot.
Thus, the system could be in one of two possible states, i.e., the states with indices $i = 0$ and $i = 1$ in Table~\ref{tbl:observation_probabilities}.
This comes in contrast to the eight possible states of the basic scenario and resulted in reduced cost due to health status entropy for all policies and the whole range of  $\tau_{SM}^f$ values.
Furthermore, in Figure~\ref{fig:scenario_1_1} we do not observe a significant reduction of $\hat{J}_0$ when $\tau_{SM}^f \geq 0.5$ as in Figure~\ref{fig:scenario_2_3}.
This is because, the decrement of health status entropy cost as $\tau_{SM}^f$ increased, in this scenario, was not as significant as the increment of normalized AoI and probing costs. 

Finally, in Figure~\ref{fig:scenario_4} we present the effect of an increasing time horizon $N$ on $\hat{J}_0(\cdot)$.
We modified the basic system setup by setting $p_{SM}^{11} = 0.9$ and $\lambda_2 = \frac{N}{100}$.
The latter modification is necessary otherwise the cost induced by the normalized AoI became negligible as the time horizon increased. 
By setting $\lambda_2 = \frac{N}{100}$ we had,
a  normalized AoI cost of $\bar{\Delta} = \frac{\Delta}{100}$, which was analogous to that of the basic scenario irrespectively to the time horizon $N$. 
Figure~\ref{fig:scenario_4} depicts that an increment of $N$ results in an increased $\hat{J}_0(\cdot)$ for all policies.
Furthermore, by calculating the relative difference of the threshold policy with respect to the delay policy with $D = 90$ we were observed that it achieved a constant reduction of $16\%$ across all experiments. 

\section{Conclusions}
In this work, we address the problem of deriving an efficient policy for sensor probing in IoT networks with intermittent faults. 
We adopted a semantics-aware communications paradigm for the transmission of probes whereby the importance (semantics) of the probe is considered before its generation and transmission.
We formulated the problem as a POMDP and proved that the optimal policy is of a threshold type and used a computationally efficient stochastic approximation algorithm to derive the probing policy. 
Finally, the numerical results presented in this work exhibit a significant cost reduction when the derived probing policy is followed instead of a conventional delay based one.

\appendices

\section{Proof of Lemma~\ref{lemma:ProbeDecrementsEntropy}} \label{apdx:ProbeDecrementsEntropy} 
First we present the part of the proof for the case where the monitor observes a fresh status update at $t+1$, i.e., $z_{t+1} = 1$. 
From Equation~(\ref{eq:NextBeliefState}) and the column of Table~\ref{tbl:observation_probabilities} that corresponds to $a_{t}=0$ and $z_{t+1} = 1$ we have that, $P_{t+1}^{0, 1}$ has only the following two non-zero elements,
$ p_{t+1}^0 = \frac{\sum_{i=0}^{7} p_t^i p_{i0} P_g}{\sum_{i=0}^{7} p_t^i(p_{i0} + p_{i4})P_g}$
and 
	$p_{t+1}^4 = \frac{\sum_{i=0}^{7} p_t^i p_{i4} P_g}{\sum_{i=0}^{7} p_t^i(p_{i0} + p_{i4})P_g}$.
We represent the numerators of $ p_{t+1}^0$ and $ p_{t+1}^4$ 
with $\phi_1 = \sum_{i=0}^{7} p_t^i p_{i0} P_g $ and $\phi_2 = \sum_{i=0}^{7} p_t^i p_{i4} P_g $ and we get, $p_{t+1}^0 = \frac{\phi_1}{\phi_1+\phi_2}$ and $p_{t+1}^4 = \frac{\phi_2}{\phi_1+\phi_2}$.
Furthermore, we can express the belief state vector at the next time slot as, 
$ P_{t+1}^{0,1} = \bigg[\frac{\phi_1}{\phi_1+\phi_2}, 0, 0, 0, \frac{\phi_2}{\phi_1+\phi_2}, 0, 0, 0\bigg]^T $
and the corresponding health status belief vector as,
$P_{t+1}^{h, 0, 1} =  \bigg[p_{t+1}^h, p_{t+1}^f\bigg] = \bigg[p_{t+1}^0 + p_{t+1}^4, 0 \bigg] = [1, 0]$.

Similarly, from Equation~(\ref{eq:NextBeliefState}) and the column of Table~\ref{tbl:observation_probabilities} that corresponds to $a_{t}=1$ and $z_{t+1} = 1$ we have that $P_{t+1}^{1, 1}$ has again only two non-zero elements,
$ p_{t+1}^0 = \frac{\sum_{i=0}^{7} p_t^i p_{i0}}{\sum_{i=0}^{7} p_t^i(p_{i0} + p_{i4}P_g)} $
and 
$ p_{t+1}^4 = \frac{\sum_{i=0}^{7} p_t^i p_{i4} P_g}{\sum_{i=0}^{7} p_t^i(p_{i0} + p_{i4}P_g)}$.
We represent the numerators of $ p_{t+1}^0$ and $ p_{t+1}^4$ with
$\xi_1 = \sum_{i=0}^{7} p_t^i p_{i0} $ and 
$\xi_2 = \sum_{i=0}^{7} p_t^i p_{i4} P_g$ and we get,
$p_{t+1}^0 = \frac{\xi_1}{\xi_1+\xi_2}$ and $p_{t+1}^4 = \frac{\xi_2}{\xi_1+\xi_2}$.

The belief state vectors at the next time slot,
will be,
$ P_{t+1}^{1,1} = \bigg[\frac{\xi_1}{\xi_1+\xi_2}, 0, 0, 0, \frac{\xi_2}{\xi_1+\xi_2}, 0, 0, 0\bigg]^T$
and the corresponding health status belief vector will be,
$	P_{t+1}^{h, 1, 1} = \bigg[p_{t+1}^0 + p_{t+1}^4, 0 \bigg] = [1, 0] $
From the expressions for $P_{t+1}^{h, 0, 1}$ and $P_{t+1}^{h, 1, 1}$ given above we get that $H(P_{t+1}^{h, 0, 1}) = H(P_{t+1}^{h, 1, 1})$. 

Next we present the case where the monitor does not observe a fresh status update at $t+1$, i.e., $z_{t+1} = 0$.
We substitute, for each state index $j$ in (\ref{eq:NextBeliefState}), the observation probabilities presented in the column of Table~\ref{tbl:observation_probabilities} that corresponds to $a_{t}=0$ and $z_{t+1} = 0$ and we derive for $j = 0$, $ p_{t+1}^0 = \frac{\sum_{i=0}^{7} p_t^i p_{i0} (1-P_g)}{\sum_{i=0}^{7} p_t^i [p_{i0}(1-P_g) + p_{i4}(1-P_g) + \sum_{j\neq0,4} p_{ij}]}$,
for $j = 4$, $ p_{t+1}^4 = \frac{\sum_{i=0}^{7} p_t^i p_{i4} (1-P_g)}{\sum_{i=0}^{7} p_t^i [p_{i0}(1-P_g) + p_{i4}(1-P_g) + \sum_{j\neq0,4} p_{ij}]}, $
and for $j = \{0, 1,\dots,7\}\setminus \{0,4\}$, $ p_{t+1}^j = \frac{\sum_{i=0}^{7} p_t^i p_{ij}}{\sum_{i=0}^{7} p_t^i [p_{i0}(1-P_g) + p_{i4}(1-P_g) + \sum_{j\neq0,4} p_{ij}]} $.
By setting $\xi_1 = \sum_{i=0}^{7} p_t^i p_{i0} (1-P_g)$, $\xi_2 = \sum_{i=0}^{7} p_t^i p_{i4} (1-P_g)$, $\phi_j = \sum_{i=0}^{7} p_t^i p_{ij}$ for $j = \{0, 1,\dots,7\}\setminus \{0,4\}$ and $\phi_s = \sum_{j\neq0,4} \phi_j$ we get that 
the belief state at $t+1$ is, $ P_{t+1}^{0, 0} = \frac{1}{\xi_1 + \xi_2 + \phi_s} \cdot \bigg[\xi_1, \phi_1, \phi_2, \phi_3, \xi_2, \phi_5, \phi_6, \phi_7 \bigg]^T.$
The resulting health status belief vector will be, $ P_{t+1}^{h, 0, 0} = \bigg[\frac{\xi_1 + \xi_2}{\xi_1 + \xi_2 + \phi_s} , \frac{\phi_s}{\xi_1 + \xi_2 + \phi_s}  \bigg]^T $.

Similarly, by substituting, for each state index $j$ in Equation~(\ref{eq:NextBeliefState}), the observation probabilities presented in the column of Table~\ref{tbl:observation_probabilities} that corresponds to $a_{t} = 1$ and $z_{t+1} = 0$, we derive for $j = 0$, $ 	p_{t+1}^0 = 0 $
for $j = 4$, $ 	p_{t+1}^4 = \frac{\sum_{i=0}^{7} p_t^i p_{i4} (1-P_g)}{\sum_{i=0}^{7} p_t^i [p_{i4}(1-P_g) + \sum_{j\neq0,4} p_{ij}]}, $
and for $j = \{0, 1,\dots,7\}\setminus \{0,4\}$, $ p_{t+1}^j = \frac{\sum_{i=0}^{7} p_t^i p_{ij}}{\sum_{i=0}^{7} p_t^i [p_{i0}(1-P_g) + p_{i4}(1-P_g) + \sum_{j\neq0,4} p_{ij}]}.$
The belief state at $t+1$ can be expressed as, $ P_{t+1}^{1, 0} = \frac{1}{\xi_1 + \xi_2 + \phi_s} \cdot \bigg[0, \phi_1, \phi_2, \phi_3, \xi_2, \phi_5, \phi_6, \phi_7 \bigg]^T. $
The resulting health status belief vector will be, $ P_{t+1}^{h, 1, 0} = \bigg[\frac{\xi_2}{\xi_2 + \phi_s} , \frac{\phi_s}{\xi_2 + \phi_s}  \bigg]^T.  $

In order to prove that $H(P_{t+1}^{h, 0, 0}) \geq H(P_{t+1}^{h, 1, 0})$ it is adequate to show that the probability distribution $P_{t+1}^{h, 0, 0}$ is closer to a uniform distribution compared to $P_{t+1}^{h, 1, 0}$, i.e., the following inequality is true, $ \bigg|\frac{\xi_1 + \xi_2}{\xi_1 + \xi_2 + \phi_s} - \frac{\phi_s}{\xi_1 + \xi_2 + \phi_s}  \bigg| \leq \bigg|\frac{\xi_2}{\xi_2 + \phi_s} - \frac{\phi_s}{\xi_2 + \phi_s}  \bigg|. $
In Appendix~\ref{apdx:xi_and_x2_leq_phi_s} we show that under  Assumption~\ref{assumption:NoObservationIndicatesFault} it holds that, $ \xi_1 + \xi_2 \leq \phi_s $ and, consequently, that $\xi_2 \leq \phi_s $, thus we have, $ \frac{\phi_s - \xi_1 + \xi_2}{\xi_1 + \xi_2 + \phi_s} \leq \frac{\phi_s - \xi_2}{\xi_2 + \phi_s}. $
With simple algebraic manipulations the equation presented above can be expressed as, $-2 \xi_1 \phi_s \leq 0$ which is true since both $\xi_1$, $\phi_s$ are probabilities.
This concludes the proof and shows that the probing action results in the same or reduced health status entropy.

\section{} \label{apdx:xi_and_x2_leq_phi_s}
In this appendix we show that Assumption~\ref{assumption:NoObservationIndicatesFault} is equivalent to $\xi_1 + \xi_2 \leq \phi_s$. 
For convenience we repeat here the definitions of $\xi_1$, $\xi_2$ and $\phi_s$,
\begin{enumerate}
	\item $\xi_1 = \sum_{i=0}^{7} p_t^i p_{i0} (1-P_g)$
	\item $\xi_2 = \sum_{i=0}^{7} p_t^i p_{i4} (1-P_g)$
	\item $\phi_s = \sum_{i=0}^{7} p_t^i \sum_{j} p_{ij}$, $j = I_S \setminus \{0,4\}$,
\end{enumerate}
where $I_S = \{0, 1,\dots,7\}$ and in $\phi_s$ we changed the order of summation.
Now, by substituting to $\xi_1 + \xi_2 \leq \phi_s$ we get, 
\begin{align} \label{eq:assumption_2_indexes}
	\sum_{i=0}^{7} p_t^i p_{i0} (1-P_g) + \sum_{i=0}^{7} p_t^i p_{i4} (1-P_g) &\leq \sum_{i=0}^{7} p_t^i \sum_{j\in I_s \setminus \{0,4\}} p_{ij} 
\end{align}
which can be written as $ \sum_{i=0}^{7} p_t^i \big[ p_{i0} (1-P_g) + p_{i4} (1-P_g) - \sum_{j\in I_s \setminus \{0,4\}} p_{ij} \big] \leq 0 $.
Subsequently we express the transition probability from state with index $i$ to state with index $j$ as $P[s_{t+1} = j|s_t = i] = p_{ij} = p_{i_0 j_0}^{MS} \cdot p_{i_1 j_1}^{S} \cdot p_{i_2 j_2}^{SM}$, where $i_0$, $i_1$, and $i_2$ represent respectively the states of the $l_{MS}$ link, the sensor and the $l_{SM}$ link at time $t$, while $j_0$, $j_1$, and $j_2$ represent respectively the states of the $l_{MS}$ link, the sensor and the $l_{SM}$ link at time $t+1$. 
Thus expression, $\big[p_{i0} (1-P_g) + p_{i4} (1-P_g) - \sum_{j\in I_s \setminus \{0,4\}} p_{ij}\big]$ 
can be written as, $\big[ p_{i_0 0}^{MS}p_{i_1 0}^{S} p_{i_2 0}^{SM} (1-P_g) + p_{i_0 1}^{MS}p_{i_1 0}^{S} p_{i_2 0}^{SM} (1-P_g) - p_{i_0 0}^{MS} p_{i_1 0}^{S} p_{i_2 1}^{SM} - p_{i_0 0}^{MS} p_{i_1 1}^{S} p_{i_2 0}^{SM} - p_{i_2 0}^{SM} p_{i_1 1}^{S} p_{i_2 1}^{SM} - p_{i_0 1}^{MS} p_{i_1 0}^{S} p_{i_2 1}^{SM} - p_{i_0 1}^{MS} p_{i_1 1}^{S} p_{i_2 0}^{SM} - p_{i_0 1}^{MS} p_{i_1 1}^{S} p_{i_2 1}^{SM} \big]$.
Through simple algebraic manipulations the latter expression can be shown to be equal to $\big[ p_{i_1 0}^{S} p_{i_2 0}^{SM} (2-P_g) -1 \big]$ and,
based on this result, (\ref{eq:assumption_2_indexes}) can be expressed as $ \sum_{i=0}^{7} p_t^i \big[ p_{i_1 0}^{S} p_{i_2 0}^{SM} (2-P_g)  - 1 \big] \leq 0. $

\section{Proof of Lemma~\ref{lemma:increasingInEntropy}} \label{apdx:increasingInEntropy}
To prove Lemma~\ref{lemma:increasingInEntropy} we will use induction.
For the $N$-th decision stage Lemma~\ref{lemma:increasingInEntropy} holds trivially, i.e., for $x_N^+ = [P_N^+, \bar{\Delta}_N]$ and $x_N^- = [P_N^-, \bar{\Delta}_N]$ with $H(P_N^{h,+}) \geq H(P_N^{h,-})$ we have, 
\begin{align*}
	H(P_N^{h,+}) + \bar{\Delta}_N \geq H(P_N^{h,-}) + \bar{\Delta}_N \Rightarrow 
	J_N(P_N^+, \Delta_N)  \geq J_N(P_N^-, \Delta_N)
\end{align*}	
Let it be true that for $x_{t+1}^+ = [P_{t+1}^+, \bar{\Delta}_{t+1}]$ and $x_{t+1}^- = [P_{t+1}^-, \bar{\Delta}_{t+1}]$ with $H(P_{t+1}^{h,+}) \geq H(P_{t+1}^{h,-})$ it holds that, $ J_{t+1}(P_{t+1}^+, \Delta_{t+1})  \geq J_{t+1}(P_{t+1}^-, \Delta_{t+1}) $
then we will prove that,
for $x_t^+ = [P_t^+, \bar{\Delta}_t]$ and $x_t^- = [P_t^-, \bar{\Delta}_t]$ with $H(P_t^{h,+}) \geq H(P_t^{h,-})$ it is also true that, 
\begin{equation} \label{eq:OptimalCostIncreasesWithEntropy}
	J_t(P_t^+, \Delta_t)  \geq J_t(P_t^-, \Delta_t)
\end{equation}	
We have, 
\begin{equation} \label{eq:OptimalCostHighEntropy}
	J_t(P_t^+, \Delta_t) = \min \big[ A_0^+, A_1^+ \big]
\end{equation}
where,
\begin{equation} \label{eq:A0+}
A_0^+ = H(P_t^{h,+}) + \bar{\Delta}_t + \sum_z \sum_s \sum_i p_t^{i,+}\, p_{is} \, r_s(0,z)\, J_{t+1}(P_{t+1}^{0, z, +}, \bar{\Delta}_{t+1}^z),
\end{equation}
\begin{equation} \label{eq:A1+}
A_1^+ = c + H(P_t^{h,+}) + \bar{\Delta}_t + \sum_z \sum_s \sum_i p_t^{i,+}\, p_{is} \, r_s(1,z)\, J_{t+1}(P_{t+1}^{1, z, +}, \bar{\Delta}_{t+1}^z)
\end{equation}
where $P_{t+1}^{0, z, +}$ ($P_{t+1}^{1, z, +}$) is the belief state calculated at the $(t+1)$-th time slot from $P_t^+$ using Equation~(\ref{eq:NextBeliefState}), when $a_t = 0$ ($a_t = 1$) and $z_{t+1}=z$.
Similarly, we have,   
\begin{equation} \label{eq:OptimalCostLowEntropy}
	J_t(P_t^-, \Delta_t) = \min \big[ A_0^-, A_1^- \big]
\end{equation}
where $A_0^-$, and $A_1^-$ are defined by replacing $+$ with $-$ in (\ref{eq:A0+}) and~(\ref{eq:A1+}) respectively.
In order to prove Equation~(\ref{eq:OptimalCostIncreasesWithEntropy}) it suffices to show that the following two inequalities hold,
\begin{align} 
	\setstretch{1.0}
	\label{eq:ProbeHiLowEntropycomparison} 	A_1^+ & \geq A_1^-, \\
	\label{eq:NoProbeHiLowEntropycomparison} A_0^+ &\geq A_0^-. 
\end{align}
This can be verified by considering all possible combinations for the values of $J_t(x_t^+)$ and $J_t(x_t^-)$ from~(\ref{eq:OptimalCostHighEntropy}) and~(\ref{eq:OptimalCostLowEntropy}):
\begin{enumerate}
\item Let $J_t(x_t^+) = A_0^+$ and $J_t(x_t^+)=A_0^-$ then by~(\ref{eq:NoProbeHiLowEntropycomparison}) Equation~(\ref{eq:OptimalCostIncreasesWithEntropy}) holds.
\item Let $J_t(x_t^+) = A_0^+$ and $J_t(x_t^+)=A_1^-$ then $A_0^+ \geq A_0^-$ by~(\ref{eq:NoProbeHiLowEntropycomparison}) and $A_0^- \geq A_1^-$ due to $A_1^-$ being optimal, i.e., due to the $\min[\cdot]$ operator in~(\ref{eq:OptimalCostLowEntropy}), thus we have $A_0^+ \geq A_1^-$ and as a result Equation~(\ref{eq:OptimalCostIncreasesWithEntropy}) holds.
\item Let $J_t(x_t^+) = A_1^+$ and $J_t(x_t^+)=A_0^-$ then $A_1^+ \geq A_1^-$ by~(\ref{eq:ProbeHiLowEntropycomparison}) and $A_1^- \geq A_0^-$ due to $A_0^-$ being optimal in~(\ref{eq:OptimalCostLowEntropy}), 
thus we have $A_1^+ \geq A_0^-$ and as a result Equation~(\ref{eq:OptimalCostIncreasesWithEntropy}) holds.
\item Finally, let $J_t(x_t^+) = A_1^+$ and $J_t(x_t^+)=A_1^-$ then by~(\ref{eq:ProbeHiLowEntropycomparison}) Equation~(\ref{eq:OptimalCostIncreasesWithEntropy}) holds.
\end{enumerate}

Next we show that inequality~(\ref{eq:ProbeHiLowEntropycomparison}) holds.
Proof that Equation~(\ref{eq:NoProbeHiLowEntropycomparison}) holds can be derived in a similar way.
Firstly, from the basic assumption of Lemma~\ref{lemma:increasingInEntropy} we have,
$	H(P_t^{h,+}) \geq H(P_t^{h,-}) \Rightarrow 
	H(P_t^{h,+}) + \bar{\Delta}_t \geq H(P_t^{h,-}) + \bar{\Delta}_t \nonumber$.
Secondly, from Assumption~\ref{assmption:causalInEntropy}, the fact that 
the value of AoI at $t+1$, i.e., $\bar{\Delta}_{t+1}^z$, will have the same value for a given observation $z_{t+1}=z$ independently of the starting belief state
and the induction hypothesis we have that, $ J_{t+1}(P_{t+1}^{1, z, +}, \bar{\Delta}_{t+1}^z) \geq J_{t+1}(P_{t+1}^{1, z, -}, \bar{\Delta}_{t+1}^z),\quad z \in \{0,1\} $.
Consequently, for each observation $z$, $J_{t+1}(P_{t+1}^{1, z, +}, \bar{\Delta}_{t+1}^z)$ is uniformly larger than $J_{t+1}(P_{t+1}^{1, z, -}, \bar{\Delta}_{t+1}^z)$ and thus its expected value at $t+1$ will also be larger~\cite[Chapter 7, pg. 299]{sheldonRossAfirstCourseInProbability}, i.e.,
\begin{equation*} \label{}
	\sum_s \sum_i p_t^{i,+}\, p_{is} \, r_s(a_t,z)\, J_{t+1}(P_{t+1}^{1, z, +}, \bar{\Delta}_{t+1}^z) \geq \sum_s \sum_i p_t^{i, -}\, p_{is} \, r_s(a_t,z)\, J_{t+1}(P_{t+1}^{1, z, -}, \bar{\Delta}_{t+1}^z) 
\end{equation*}
and by summing over all possible observations we get,
\begin{equation*} 
	\sum_z \sum_s \sum_i p_t^{i,+}\, p_{is} \, r_s(a_t,z)\, J_{t+1}(P_{t+1}^{1, z, +}, \bar{\Delta}_{t+1}^z) \geq \sum_z \sum_s \sum_i p_t^{i, -}\, p_{is} \, r_s(a_t,z)\, J_{t+1}(P_{t+1}^{1, z, -}, \bar{\Delta}_{t+1}^z)
\end{equation*}
which concludes the proof.

\section{Proof of Lemma~\ref{lemma:linearJinEntropy}} \label{apdx:linearJinEntropy}
We prove Lemma~\ref{lemma:linearJinEntropy} using induction. 
At the final stage, $t = N$, we have that $J_N(P,\bar{\Delta}) = \lambda_1 H(P^h) + \lambda_2 \bar{\Delta}$ which is linear in $H(P^h)$ and $\bar{\Delta}$.
For stage $t = N-1$, we have that, 
\begin{equation} \label{eq:OptimalCostToGoN}
	\setstretch{1.0}\
	J_{N-1}(x_{N-1}) = \min \big[ A_{0, N-1}, A_{1, N-1} \big]
	\setstretch{1.0}\
\end{equation}
and
\begin{equation} \label{eq:A0N}
	\setstretch{1.0}\
	A_{0, N-1} = \lambda_1H(P_{N-1}^h) + \lambda_2\bar{\Delta}_{N-1} + \sum_z \sum_s \sum_i p_{N-1}^{i}\, p_{is} \, r_s(0,z)\, J_{N}(P_{N}^{0, z}, \bar{\Delta}_{N}^z),
\end{equation}
\begin{equation} \label{eq:A1N}
	A_{1, N-1} = c + \lambda_1 H(P_{N-1}^h) + \lambda_2\bar{\Delta}_{N-1} + \sum_z \sum_s \sum_i p_{N-1}^{i}\, p_{is} \, r_s(1,z)\, J_{N}(P_{N}^{1, z}, \bar{\Delta}_{N}^z)
\end{equation}
By the minimization operator in Equation~(\ref{eq:OptimalCostToGoN}), the fact that $A_{0, N-1}$ and $A_{1, N-1}$ are linear and increasing in $H(P)$ and $\bar{\Delta}$ we have $J_{N-1}(x)$ is piece-wise linear, increasing and concave with respect to $H(P)$ and $\bar{\Delta}$.
Assuming that $J_{t+1}(P_{t+1}, \bar{\Delta}_{t+1})$ is piece-wise linear, increasing and concave with respect to $H(P_{t+1})$ and $\bar{\Delta}_{t+1}$ then we will show that $J_t(P_t, \bar{\Delta}_t)$ will also be piece-wise linear, increasing and concave with respect to $H(P_{t+1})$ and $\bar{\Delta}_{t+1}$.
For $J_t(P_t, \bar{\Delta}_t)$ we have,
\begin{equation} \label{eq:OptimalCostToGo}
	\setstretch{1.0}\
	J_t(x_{t}) = \min \big[ A_0, A_1 \big]
\end{equation}
and
\begin{equation} \label{eq:A0}
	\setstretch{1.0}\
	A_{0, t} = \lambda_1H(P_t^h) + \lambda_2\bar{\Delta}_t + \sum_z \sum_s \sum_i p_t^{i}\, p_{is} \, r_s(0,z)\, J_{t+1}(P_{t+1}^{0, z}, \bar{\Delta}_{t+1}^z),
\end{equation}
\begin{equation} \label{eq:A1}
	\setstretch{1.0}\
	A_{1,t}= c + \lambda_1 H(P_t^h) + \lambda_2\bar{\Delta}_t + \sum_z \sum_s \sum_i p_t^{i}\, p_{is} \, r_s(1,z)\, J_{t+1}(P_{t+1}^{1, z}, \bar{\Delta}_{t+1}^z).
\end{equation}
From (\ref{eq:A0}) and~(\ref{eq:A1}) and the induction hypothesis we have that $A_{0,t}$ and $A_{1,t}$ are piece-wise linear, increasing and concave with respect to $H(P_{t+1}^h)$ and $\bar{\Delta}_{t+1}$. 
Finally, considering (\ref{eq:OptimalCostToGo}) we have that $J_t(x)$ is piece-wise linear, increasing and concave with respect to $H(P^h)$ and $\bar{\Delta}$ for $t= 1,\dots, N$.

\section{Proof of Theorem~\ref{thm:monotonePolicy}} \label{apdx:monotonePolicy}
In order to prove Theorem~\ref{thm:monotonePolicy} we will show that there exist  positive probing cost values $c$ such that the optimal action at decision stage $t = N-1, \dots, 1, 0$ is the probe action.
Subsequently we will show that, if the optimal action is the probe action for a given state $x_t = [P_t, \bar{\Delta}_t]$ and cost value $c$ then the optimal action will be the probe action for all states $x_t$ with higher health status entropy. 
With similar arguments it can be shown that the probe action will also be optimal for states with higher normalized AoI than $x_t$.
As a result, at each decision stage $t$, the optimal policy will be threshold based with respect to $V_t$.

The Bellman Equation~(\ref{eq:dynamicProgram}), 
indicates that the probe action will be optimal if the following inequality is true,
\begin{multline}
	\setstretch{1.0}\
	\label{eq:ProbeInequality}
c + V_t + \sum_{z,s,i} p_t^i\, p_{is} \, r_s(1,z)\, J_{t+1}(P_{t+1}^{1,z}, \bar{\Delta}_{t+1}^z) \leq
 V_t + \sum_{z,s,i} p_t^i\, p_{is} \, r_s(0,z)\, J_{t+1}(P_{t+1}^{0,z}, \bar{\Delta}_{t+1}^z)
\end{multline}
where we have substituted $x_{t+1}$ with its constituent elements $P_{t+1}$ and $\bar{\Delta}_{t+1}$.
The superscripts $a$, $z$ in $P_{t+1}^{a,z}$ present action $a_t$ and observation $z_{t+1}$, respectively, which determine the evolution of the belief state from time slot $t$ to time slot $t+1$. 
Similarly, the superscript $z$ in $\bar{\Delta}_{t+1}^z$ presents the observation $z_{t+1}$ that determines the evolution of the normalized AoI from time slot $t$ to time slot $t+1$.
With simple algebraic manipulations Equation~(\ref{eq:ProbeInequality}) can be written as,
\begin{multline*}
	\setstretch{1.0}\
	c  \leq \sum_s \sum_i p_t^i\, p_{is} \, \big[ r_s(0,0)\,  J_{t+1}(P_{t+1}^{0,0}, \bar{\Delta}_{t+1}^0) - r_s(1,0)\, J_{t+1}(P_{t+1}^{1,0}, \bar{\Delta}_{t+1}^0)\big]+ \\ 
	 \sum_s \sum_i p_t^i\, p_{is} \, \big[ r_s(0,1)\,  J_{t+1}(P_{t+1}^{0,1}, \bar{\Delta}_{t+1}^1) - r_s(1,1)\, J_{t+1}(P_{t+1}^{1,1}, \bar{\Delta}_{t+1}^1)\big].
\end{multline*}
Substituting $r_s(a, z)$ for each combination of $s$, $a$ and $z$ as presented in Table~\ref{tbl:observation_probabilities} we get, 
\begin{multline} \label{eq:thresholdexistence}
	c \leq \sum_i p_t^i \bigg[p_{i0} \big[(1-P_g) J_{t+1}(P_{t+1}^{0,0}, \bar{\Delta}_{t+1}^0)\big]  + 
	   p_{i4} (1-P_g) \big[ J_{t+1}(P_{t+1}^{0,0}, \bar{\Delta}_{t+1}^0) - J_{t+1}(P_{t+1}^{1,0}, \bar{\Delta}_{t+1}^{0})\big] + \\
		+ p_{i0} \big[P_g J_{t+1}(P_{t+1}^{0,1}, \bar{\Delta}_{t+1}^1) - J_{t+1}(P_{t+1}^{1,1}, \bar{\Delta}_{t+1}^1) \big] + 
	    p_{i4} P_g \big[ J_{t+1}(P_{t+1}^{0,1}, \bar{\Delta}_{t+1}^1) - J_{t+1}(P_{t+1}^{1,1}, \bar{\Delta}_{t+1}^{1})\big] +\\
	    + \sum_{s \in S \setminus \{0,4\}} p_{is} \big[J_{t+1}(P_{t+1}^{0,0}, \bar{\Delta}_{t+1}^0) - J_{t+1}(P_{t+1}^{1,0}, \bar{\Delta}_{t+1}^{0})\big]  \bigg],
\end{multline}
where $S=\{0, 1, \cdots, 7\}$. 
If the right hand side of (\ref{eq:thresholdexistence}) is greater than zero then there exists a probing cost $c$ such that the optimal action is the probe action.
Given that the transition cost is positive for all decision stages, i.e., $g_t > 0$ for $t = N-1, \dots, 0, 1$, we have that the minimum expected cost from decision stage $t+1$ up to the last decision stage $N-1$ will be greater than $0$, i.e., $J_{t+1}(P_{t+1}^{0,0}, \bar{\Delta}_{t+1}^0) > 0$ and consequently the first term in~(\ref{eq:thresholdexistence}) is strictly positive. 
It remains to show that, $ J_{t+1}(P_{t+1}^{0,0}, \bar{\Delta}_{t+1}^0) - J_{t+1}(P_{t+1}^{1,0}, \bar{\Delta}_{t+1}^{0}) \geq 0 $ and $ J_{t+1}(P_{t+1}^{0,1}, \bar{\Delta}_{t+1}^1) - J_{t+1}(P_{t+1}^{1,1}, \bar{\Delta}_{t+1}^{1}) \geq 0 $.
From Lemma~\ref{lemma:ProbeDecrementsEntropy} we have that $H(P_{t+1}^{h,0,0}) \geq H(P_{t+1}^{h,1,0}) $ and $H(P_{t+1}^{h,0,1}) \geq H(P_{t+1}^{h,1,1})$ and according to Lemma~\ref{lemma:increasingInEntropy} we have that $J_{k}(P, \bar{\Delta})$ is an increasing function of the health status entropy for all decision stages, thus the above inequalities
are true. 

Now assume that the probe action is optimal for the state $x_t = [P_t, \bar{\Delta}_t]$ and let $x_t^{+} = [P_t^+, \bar{\Delta}_t]$ be such that $H(P_t^{h,+}) \geq H(P_t^h)$.
Furthermore, let $P_{t+1}^{a, z, +}$ be the belief state at $t+1$ given that the belief state at $t$ was $P_t^+$, action $a$ was taken at time $t$ and observation $z$ was made at $t+1$. 
From Lemma~\ref{lemma:ProbeDecrementsEntropy} we have that probing results in a reduction of entropy, i.e., $H(P_{t+1}^{h,0,0,+}) \geq  H(P_{t+1}^{h,1,0,+})$ and $H(P_{t+1}^{h,0,1,+}) \geq H(P_{t+1}^{h,1,1,+})$, from Lemma~\ref{lemma:increasingInEntropy} and Lemma~\ref{lemma:linearJinEntropy} we have that $J_t(\cdot)$, $t = 0,1,\dots, N$ is piece-wise linear, increasing and concave with respect entropy, thus the following inequalities,
\begin{enumerate}
	\item $ J_{t+1}(P_{t+1}^{0,0, +}, \bar{\Delta}_{t+1}^0) \geq 0 $
	\item $ J_{t+1}(P_{t+1}^{0,0, +}, \bar{\Delta}_{t+1}^0) - J_{t+1}(P_{t+1}^{1,0, +}, \bar{\Delta}_{t+1}^{0}) \geq 0 $
	\item $ J_{t+1}(P_{t+1}^{0,1, +}, \bar{\Delta}_{t+1}^1) - J_{t+1}(P_{t+1}^{1,1, +}, \bar{\Delta}_{t+1}^{1}) \geq 0 $
\end{enumerate}
hold. As a result there exists a probing cost $c > 0$ such that the probing action is optimal for $x_t^+$ as well.
What is more, according to Lemmas~\ref{lemma:increasingInAoI} and~\ref{lemma:linearJinEntropy}, $J_t(\cdot)$ is also increasing in $\bar{\Delta}_t$ and, as a consequence, it will also be increasing when both $\bar{\Delta}_t$ and $P_t$ increase. 
Thus, the same proposition as above will be true for all states $x_t^{+} = [P_t^{h,+}, \bar{\Delta}_t^+]$ such that $H(P_t^{h,+}) \geq H(P_t^h)$ and $\bar{\Delta}_t^+ \geq \bar{\Delta}_t$.

\bibliographystyle{IEEEtranTCOM}
\bibliography{IEEEabrv, bibliography}

\end{document}